\documentclass[prints]{aa}
\usepackage[authoryear]{natbib}
\bibpunct[, ]{(}{)}{;}{a}{}{,}
\usepackage{graphicx}

\begin{document}

\title{Hubble Space Telescope Imaging of Globular Cluster Candidates in Low
Surface Brightness Dwarf Galaxies\thanks{Based on observations made with
the NASA/ESA Hubble Space Telescope. The Space Telescope Science Institute
is operated by the Association of Universities for Research in Astronomy,
Inc. under NASA contract NAS 5--26555.}}
\titlerunning{HST Imaging of Globular Cluster Candidates in nearby LSB dwarf galaxies}
\author{
     M. E. Sharina \inst{1,2}
\and T. H. Puzia   \inst{3,}\thanks{ESA Fellow, Space Telescope Division of ESA}
\and D. I. Makarov \inst{1,2}}
\institute{Special Astrophysical Observatory, Russian Academy
of Sciences, N. Arkhyz, KChR, 369167, Russia
\and Isaac Newton Institute, Chile, SAO Branch
\and Space Telescope Science Institute, 3700 San Martin Drive, Baltimore, MD21218, USA}
\date{Received: February 2005 -- Accepted: May 2005}
\abstract{
Fifty-seven nearby low surface brightness dwarf galaxies ($-10\ga M_{V}\ga
-16$) were searched for globular cluster candidates (GCCs) using {\it
Hubble Space Telescope} WFPC2 imaging in $V$ and $I$. The sample consists
of 18 dwarf spheroidal (dSph), 36 irregular (dIrr), and 3 "transition"
type (dIrr/dSph) galaxies with angular sizes less than 3.7 kpc situated at
distances $2\!-\!6$ Mpc in the field and in the nearby groups: M81,
Centaurus A, Sculptor, Canes Venatici~I cloud. We find that $\sim50$\% of
dSph, dIrr/dSph, and dIrr galaxies contain GCCs. The fraction of GCCs
located near the center of dwarf spheroidal galaxies is $\ga2$ times
higher than that for dIrrs. The mean integral color of GCCs in dSphs,
$(V-I)_0=1.04\pm0.16$ mag, coincides with the corresponding value for
Galactic globular clusters  and is similar to the {\it blue} globular
cluster sub-populations in massive early-type galaxies. The color
distribution for GCCs in dIrrs shows a clear bimodality with peaks near
$(V-I)_0 = 0.5$ and $1.0$ mag. Blue GCCs are presumably young with
ages $t\la1$ Gyr, while the red GCC population is likely to be older. The
detected GCCs have absolute visual magnitudes between $M_{V}=-10$ and $-5$
mag. We find indications for an excess population of faint GCCs with
$M_{V}\ga-6.5$ mag in both dSph and dIrr galaxies, reminiscent of excess
populations of faint globular clusters in nearby Local Group spiral
galaxies. The measurement of structural parameters using King-profile
fitting reveals that most GCCs have structural parameters similar to
extended outer halo globular clusters in the Milky Way and M31, as
well as the recently discovered population of ''faint fuzzy'' clusters in
nearby lenticular galaxies.
}
\maketitle


\section{Introduction}
Low surface brightness dwarf galaxies ($M_V\! >\!-16$, and central
surface brightness $\mu_{\rm V}\!\ga\!22$ mag/arcsec$^2$) constitute the
most numerous galaxy type in the local universe. Their formation
mechanisms, their physical structure, and their contribution to the
assembly of massive galaxies has attracted great attention since many
years \citep[e.g.][]{ferguson94, impey97, bothun97, klypin99, kravtsov04}.
Star formation in these low-mass stellar systems appears to be governed by
a complex interplay of gravitational instabilities, turbulence, and gas
thermodynamics \citep[e.g.][]{elmegreen02, pelupessy04}. In extreme
cases, star formation culminates in the formation of massive star clusters
that are likely to be the young counterparts of today's old globular
clusters \citep[e.g.][]{larsen00}. This mode of star-formation is observed
in numerous nearby dwarf irregular galaxies \citep[dIrr;
e.g.][]{billett02, hunter04}, but appears to have ceased a long time ago
in dwarf spheroidal (dSph) and dwarf elliptical galaxies \citep[dE;
see][]{lotz04}\footnote{We explicitly consider in our sample the type of
tidal dwarf galaxies \citep[e.g.][]{weilbacher02}.}. While dE and dSph
galaxies are predominantly found in the vicinity of massive galaxies in
galaxy groups and rich galaxy clusters, dIrr galaxies are predominantly
located in the field and in loose groups. The conditions for globular
cluster formation in dwarf galaxies at a given galaxy mass might be
therefore a sensitive function of environmental density. It is still not
clear whether dE, dSph, and dIrr share similar formation and/or early
evolution histories \citep{davies88, marlowe99}. However, the three galaxy
types share one property: they all harbor globular clusters that are older
than several Gyr, which indicates that at least early globular cluster
formation took place irrespective of the morphological type. One can use
these globular clusters to investigate the early star formation and
chemical evolution histories of these galaxies. Because they are composed
of stars of one age and chemical composition, globular clusters offer a
unique tool to access the star formation histories of individual galaxies
\citep[e.g.][]{az98, kp00, harris01}.

In the present epoch, the formation of massive bound star clusters seems to
be associated with high-pressure environment and powerful star formation
events \citep[e.g.][]{elmegreen97, ashman01}. High-pressure environment naturally
occurs in dwarf galaxies due to their low metallicities and high critical
densities for star formation. Since globular clusters sample the chemical
conditions during major star formation events in a galaxy, the chemical
composition of globular clusters in dE, dSph, and dIrr galaxies might
provide crucial information on star formation histories and 
mechanisms that are important inputs for hierarchical galaxy formation
models. We therefore embarked on a spectroscopic survey of globular
clusters in nearby low surface brightness galaxies in and outside the
Local Group (LG).

In this paper, we present the photometric study of globular cluster
candidates. In Section~\ref{ln:obsred} we describe the galaxy sample and
data reduction steps. Section~3 deals with the cluster candidate
selection, while in Section~4 we investigate the properties of the
globular cluster candidates (GCCs).

\section{Observations and Data Reduction}
\label{ln:obsred}

\begin{figure*}[!t]
\centering
\includegraphics[width=8.5cm]{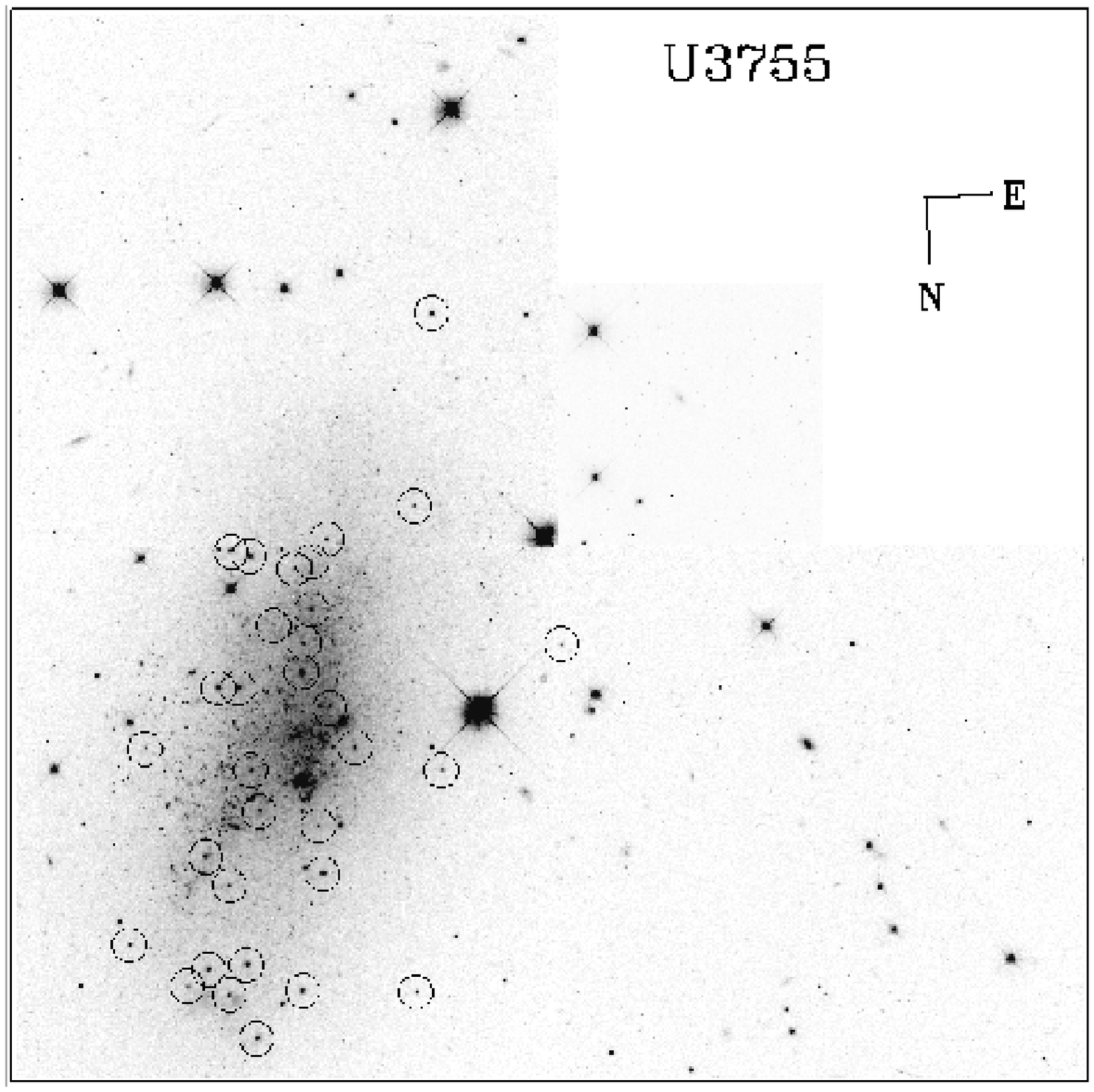}
\includegraphics[width=8.5cm]{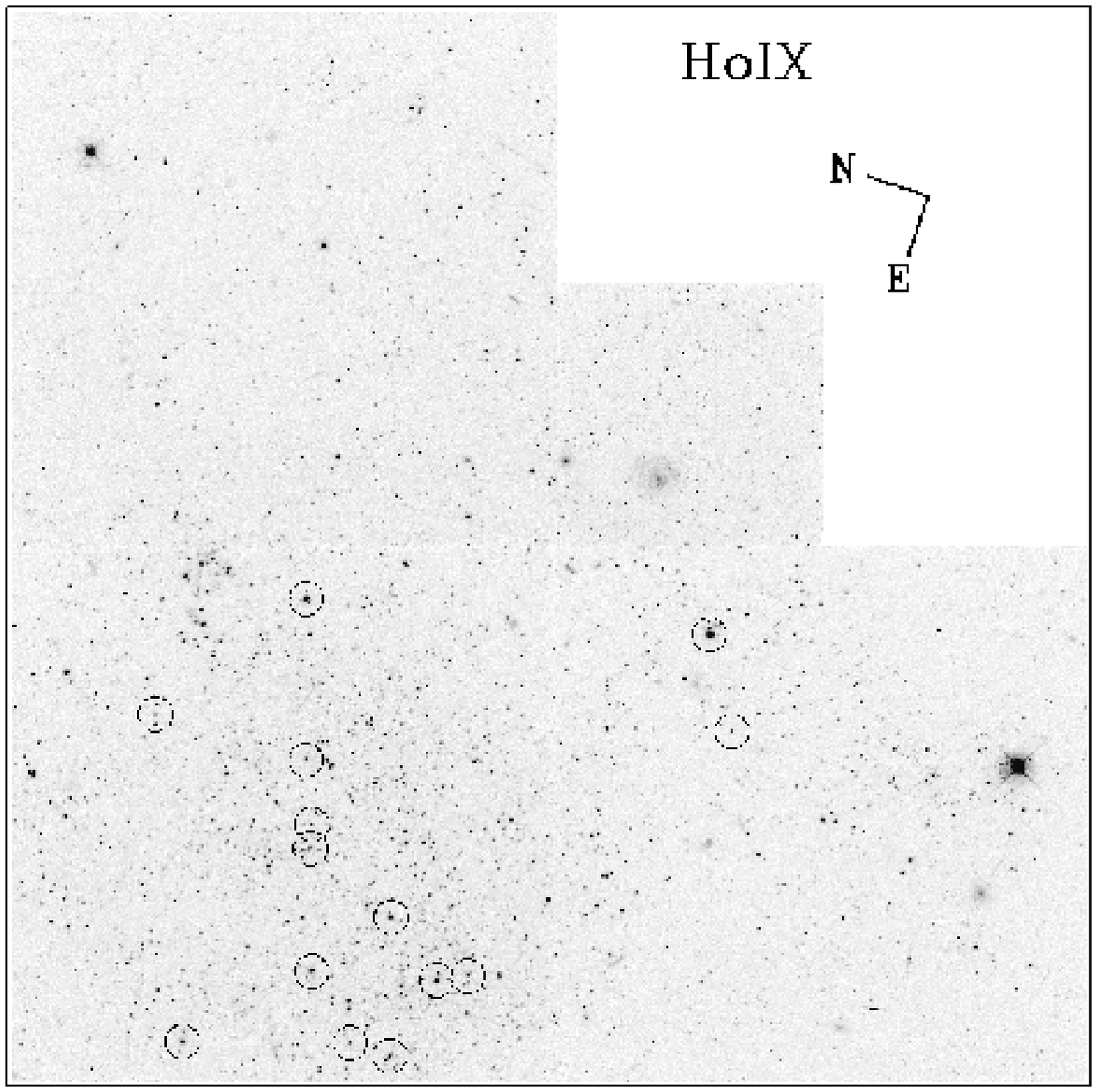}
\includegraphics[width=8.5cm]{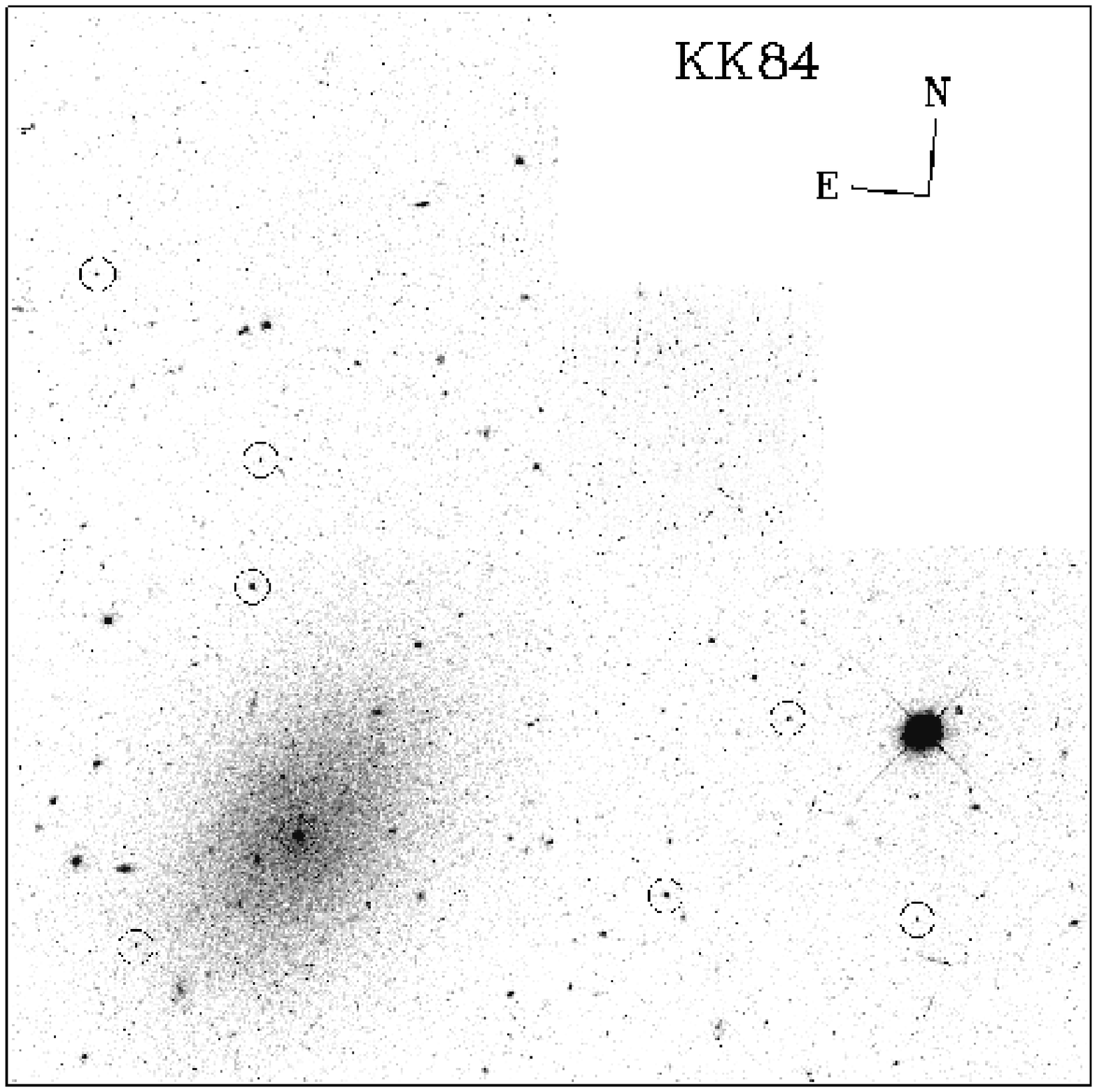}
\includegraphics[width=8.5cm]{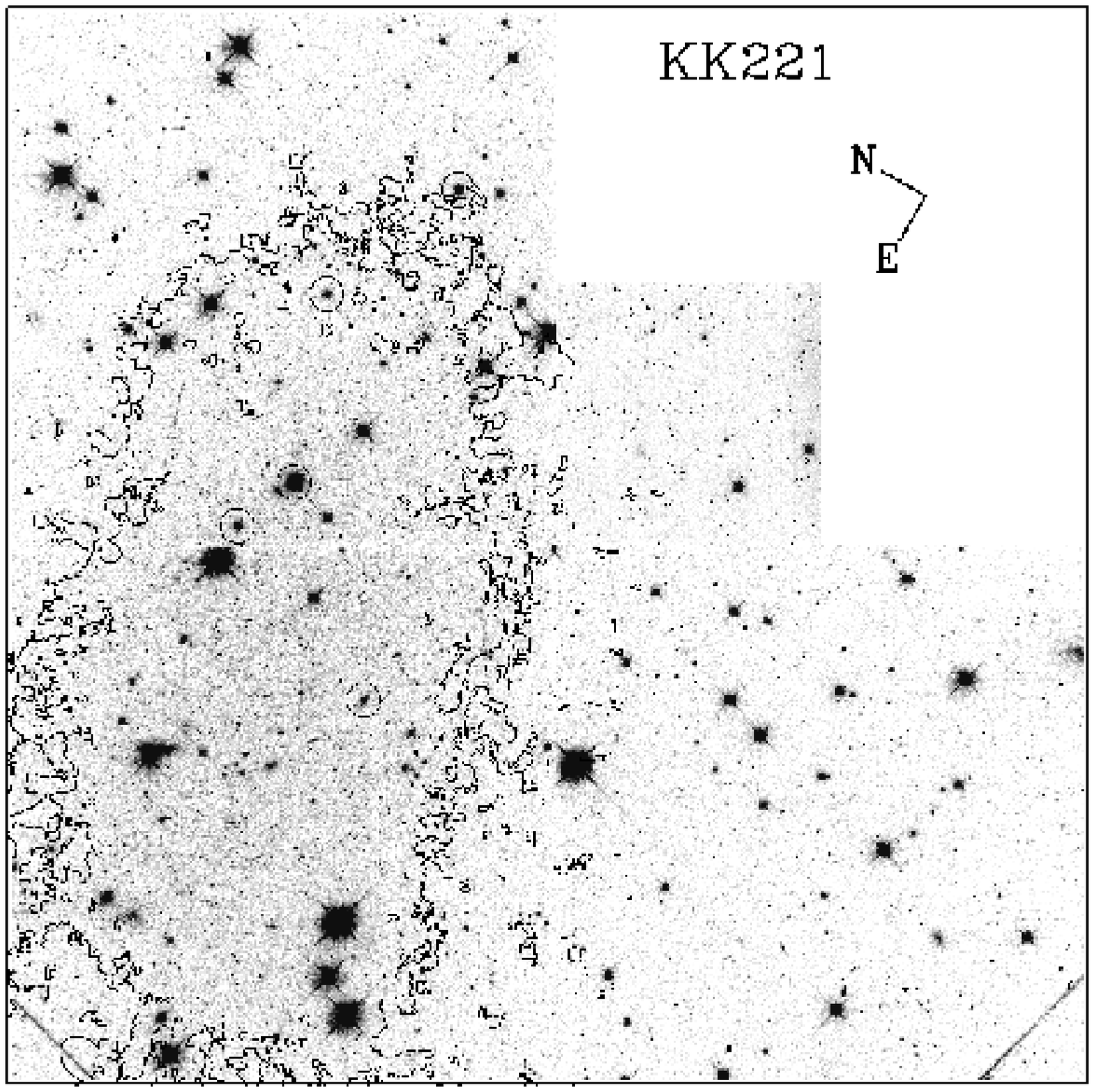}
\caption{HST/WFPC2 images for U3755, Holmberg IX, KK84, and KK221 which host 
the richest GCC systems in our sample in their morphology classes 
(see Section~1 for details).}
\label{ps:image}
\end{figure*}

Before discussing the properties of globular cluster candidates in nearby
dwarf galaxies, we briefly summarize the main characteristics of the
objects of interest. Globular clusters are centrally concentrated, mostly
spherical systems with masses of $10^4\!\le\!M_{\odot}\!\le\!10^{6.6}$,
luminosities from $M_{V}=-10.55$ (Mayall~II in M31) to $+0.2$ mag (AM--4
in the Milky Way), and half-light radii ranging from $\sim\!0.3$ to
$\sim\!25$ pc. They are bound objects whose lifetimes may exceed a Hubble
time. So far, globular clusters have been detected in 12 of the 36 Local
Group galaxies \citep{hodge02}. Table~1 presents a list of galaxies that
were searched for globular cluster candidates in this study.

Numerical values in the columns 3--4 and 6--8 were extracted and/or
computed from data presented in \cite{kara04}. Surveying Table~1 shows
that our sample is composed of 18 dwarf spheroidal (dSph: $T<-1$), 36
dwarf irregular (dIrr: $ T > 9$), and 3 intermediate-type (dSph/dIrr:
$T=-1$) galaxies with mean surface brightnesses $\mu_B > 23$
mag/arcsec$^{2}$ and angular sizes less than 3.7 kpc. All galaxies, except
KK84\footnote{KK84 is located at a distance of 9.7 Mpc.}, are situated at
distances $\sim 2 - 6$ Mpc in the field and in the nearby groups: M81,
Centaurus A, Sculptor, Canes Venatici~I cloud (see Tab.~\ref{tab:prop} for
details). We find globular cluster candidates in 10 of 18 dSphs, 18 of 36
dIrrs, and 2 of 3 intermediate-type dwarfs. In general, roughly 50\% of
all surveyed galaxies contain globular cluster candidates, irrespective of
morphological type.

\begin{table*}[!p]
\center
\label{tab:prop}
\caption{Low surface brightness galaxies in nearby groups and in the field, which
were searched for globular cluster candidates. Columns of the table
contain the following data: (1) Galaxy Name, (2) equatorial coordinates (J2000),
(3) morphological type according to RC3 \citep{RC3}, (4) distance in Mpc, (5) integrated 
absolute $V$ magnitude (indices refer to 0: this work; K0: \cite{kara0}; K1a: \cite{kara01a};
K1b: \cite{kara01b}; K1c: \cite{kara01c}; RC3: \cite{RC3}; M98: \cite{mak98}; M99: 
\cite{mak99}), (6) logarithmic surface gas density of neutral hydrogen in 
$M_{\sun}$/kpc$^2$, (7) semi-major axis diameter in kpc, (8) mean surface brightness 
in $B$-band, (9) number of globular cluster candidates, including (10) number of GCCs 
located outside the isophote of constant surface brightness $\mu_B\!=\!26.5$ mag/arcsec$^2$.}
\scriptsize
\begin{tabular}{lcrrrlrrrrlc} \\ \hline\hline\noalign{\smallskip}
Name            & RA (2000)& DEC (2000) & T & Dist & $ M_V $   & $log \Sigma_{HI}$ & $R_{\rm kpc}$ & $SB_{\rm mean}$ & $ N_{\rm GCC}$ & $ N_{\rm out}$ \\\noalign{\smallskip}
\hline                                                                                                                           \\
{\bf M81 group} &          &           &    &       &                   &       &     &      &    &      \\
KDG52           & 08 23 56 & +71 01 46 & 10 &  3.55 & $-11.71_0$        & 7.0   & 1.3 & 25.5 &  0 & 0    \\
DDO53           & 08 34 07 & +66 10 45 & 10 &  3.56 & $-13.74_{RC3}$    & 7.3   & 1.7 & 24.1 &  1 & 0    \\
A0952+69        & 09 57 29 & +69 16 20 & 10 &  3.87 & $-11.84_0$        &\dots & 2.0 & 26.5 &  0 & 0    \\
BK3N            & 09 53 49 & +68 58 09 & 10 &  4.02 & $-9.59_0$         &\dots & 0.5 & 25.6 &  1 & 1    \\
KDG73           & 10 52 55 & +69 32 45 & 10 &  3.70 & $-11.31_0$        & 7.0   & 0.6 & 24.5 &  1 & 1    \\
FM1             & 09 45 10 & +68 45 54 & $-$3 &  3.42 & $-11.04_{K1c}$ &\dots & 0.9 & 25.7 &  0 & 0    \\
KK77            & 09 50 10 & +67 30 24 & $-$3 &  3.48 & $-12.21_{K1a}$  &\dots & 2.4 & 26.4 &  3 & 0    \\
KDG61,KK81      & 09 57 03 & +68 35 30 & $-$1 &  3.60 & $-13.58_{K1a}$  &\dots & 2.5 & 25.4 &  1 & 0    \\
KKH57,HS108     & 10 00 16 & +63 11 06 & $-$3 &  3.93 & $-10.90_{K1c}$ &\dots & 0.7 & 25.3 &  0 & 0    \\
KDG63,KK83      & 10 05 07 & +66 33 18 & $-$3 &  3.50 & $-12.82_{K1a}$  &\dots & 1.7 & 25.5 &  1 & 0    \\
KDG64,KK85      & 10 07 02 & +67 49 39 & $-$3 &  3.70 & $-13.24_{K1a}$  &\dots & 2.0 & 25.1 &  0 & 0    \\
DDO78,KK89      & 10 26 28 & +67 39 24 & $-$3 &  3.72 & $-12.75_{K1a}$  &\dots & 2.1 & 25.8 &  2 & 0    \\
BK6N,KK91       & 10 34 32 & +66 00 42 & $-$3 &  3.85 & $-11.93_{K1a}$  &\dots & 1.2 & 25.5 &  2 & 0    \\
Garland         & 10 03 42 & +68 41 36 & 10 & 3.7   & $ \dots$          &  7.3  & 4.3 & \dots  &1 & \dots \\
Holmberg IX     & 09 57 32 & +69 02 35 & 10 & 3.7   & $-13.8_0$         &  7.9  & 2.5 & 24.8 &  14& 0    \\
		&          &           &    &       &                   &       &     &      &    &      \\
{\bf Sculptor group}&      &           &    &       &                   &       &     &      &    &      \\
E410$-$005,KK3  & 00 15 31 & $-$32 10 48 & $-$1 & 1.92 & $-12.11_{K0}$ &\dots & 0.7 & 24.0 &  0 & 0    \\
KDG2,E540$-$030   & 00 49 21 & $-$18 04 28 & $-$1 & 3.40 & $-12.00_{0}$ &\dots & 1.2 & 25.3 &  1 & 1    \\
E294$-$010,PGC1641& 00 26 33 & $-$41 51 20 & $-$3 & 1.92 & $-11.40_{0}$ &\dots & 0.6 & 24.2 &  1 & 0    \\
E540$-$032,FG24   & 00 50 25 & $-$19 54 25 & $-$3 & 3.42 & $-11.84_{0}$ &\dots & 1.3 & 25.6 &  0 & 0    \\
KK27            & 03 21 06 & $-$66 19 22 & $-$3  &  3.98 & $-12.32_{0}$ &\dots & 1.4 & 24.9 &  1 & 0    \\
Sc22            & 00 23 52 & $-$24 42 18 & $-$3  &  4.21 & $-11.10_{0}$ &\dots & 1.1 & 26.0 &  3 & 2    \\
DDO6            & 00 49 49 & $-$21 00 58 & 10   & 3.34 & $-12.92_{RC3}$ &  6.8  & 1.7 & 24.5 &  0 & 0    \\
		&          &              &     &       &               &       &     &      &    &      \\
{\bf CVn I cloud}   &      &              &     &       &               &       &     &      &    &      \\
KK166           & 12 49 13 & +35 36 45 & $-$3 &  4.74 & $-11.29_{0}$    &\dots & 1.0 & 25.2 &  0 & 0    \\
DDO113,KDG90    & 12 14 58 & +36 13 08 & 10   &  2.86 & $-12.67_{0}$    &\dots & 1.2 & 25.1 &  2 & 1    \\
U7605           & 12 28 39 & +35 43 05 & 10 &  4.43 & $-13.88_{M98}$    & 7.3   & 1.4 & 23.4 &  1 & 0    \\
KK109           & 11 47 11 & +43 40 19 & 10 &  4.51 & $-10.19_{0} $     & 6.9   & 0.8 & 25.9 &  1 & 1    \\
U7298           & 12 16 29 & +52 13 38 & 10 &  4.21 & $-12.54_{M99}$    & 7.2   & 1.3 & 24.5 &  1 & 0    \\
U8308,DDO167    & 13 13 22 & +46 19 18 & 10 &  4.19 & $-12.48_{M99}$    & 7.2   & 1.3 & 24.0 &  4 & 2    \\
U8833           & 13 54 49 & +35 50 15 & 10 &  3.19 & $-12.73_{M98}$    & 7.4   & 0.8 & 23.5 &  0 & 0    \\
		&          &           &    &       &                   &       &        &   &    &      \\
{\bf Cent A group}&        &           &    &       &                   &       &        &   &    &      \\
KK211           & 13 42 06 & $-$45 12 18 & $-$5 & 3.58 & $-12.58_{0}$   &\dots & 0.8 & 24.1 &  2 & 0    \\
KK213           & 13 43 36 & $-$43 46 09 & $-$3 & 3.63 & $-11.12_{0} $  &\dots & 0.6 & 25.5 &  0 & 0    \\
KK217           & 13 46 17 & $-$45 41 05 & $-$3 & 3.84 & $-12.18_{0}$   &\dots & 0.7 & 24.7 &  0 & 0    \\
KK221           & 13 48 46 & $-$46 59 49 & $-$3 & 3.98 & $-11.96_{0}$   &\dots & 1.7 & 27.0 &  5 & 0    \\
E269$-$37,KK179  & 13 03 34 & $-$46 35 03 & $-$3 & 3.48 & $-12.57_{0}$  &\dots & 0.8 & 24.0 &  0 & 0    \\
KK200           & 13 24 36 & $-$30 58 20 &  9 &  4.63 &   $-12.74_{0}$  & 6.6   & 1.8 & 25.5 &  1 & 0    \\
E444$-$84         & 13 37 20 & $-$28 02 46 & 10 &  4.61 & $-13.63_{0}$  & 7.6   & 1.7 & 24.0 &  0 & 0    \\
		&          &           &    &    &      &               &       &      &     &    &      \\
{\bf N3115 group} &        &           &    &    &      &               &       &      &     &    &      \\
KK84            & 10 05 34 & $-$07 44 57 & $-$3 & 9.7   & $-14.40_{0}$  & \dots  & 4.1 & 25.4 &  8 & 2    \\
		&          &           &    &    &      &               &       &      &     &    &      \\
{\bf Field}     &          &           &    &    &      &               &       &      &     &    &      \\
KKR25           & 16 13 48 & +54 22 16 & 10 &  1.86    & $-10.45_{K1b}$ & 6.7  & 0.6 & 25.0 &  0 & 0    \\
U8508           & 13 30 44 & +54 54 36 & 10 &  2.56   & $-13.42_{M99}$  & 7.3   & 1.3 & 23.6 &  0 & 0    \\
DDO190,U9240    & 14 24 44 & +44 31 33 & 10 &  2.79 &   $-14.37_{RC3}$  & 7.4   & 1.5 & 22.9 &  1 & 0    \\
E379$-$07,KK112   & 11 54 43 & $-$33 33 29 & 10 &  5.22 & $-12.28_{0}$  & 6.9   & 1.7 & 25.1 &  3 & 0    \\
E321$-$014        & 12 13 50 & $-$38 13 53 & 10 &  3.19 & $-13.18_{0}$  & 6.8   & 1.3 & 24.0 &  0 & 0    \\
KKH5            & 01 07 33 & +51 26 25 & 10 &  4.26 &     $-12.64_{0}$  & 7.1   & 0.7 & 23.7 &  0 & 0    \\
KKH34,Mai13     & 05 59 41 & +73 25 39 & 10 &  4.61 &     $-12.70_{0}$  & 6.9   & 1.2 & 24.7 &  0 & 0    \\
KKH98           & 23 45 34 & +38 43 04 & 10 &  2.45 &     $-11.28_{0}$  & 7.1   & 0.7 & 25.0 &  0 & 0    \\
KK16            & 01 55 21 & +27 57 15 & 10 &  4.74 &     $-12.81_{0}$  & 6.7   & 1.1 & 23.9 &  0 & 0    \\
KK17            & 02 00 10 & +28 49 57 & 10 &  4.72 &     $-11.95_{0}$  & 6.7   & 0.8 & 24.3 &  0 & 0    \\
KKH18           & 03 03 06 & +33 41 40 & 10 &  4.43 &     $-12.84_{0}$  & 7.3   & 0.9 & 23.8 &  0 & 0    \\
E489$-$56,KK54    & 06 26 17 & $-$26 15 56 & 10 & 4.99 &  $-13.51_{0}$  & 7.4   & 0.9 & 22.8 &  0 & 0    \\
E490$-$17,PGC19337 & 06 37 57 & $-$25 59 59 & 10 & 4.23 & $-14.91_{0}$  & 6.9   & 2.1 & 23.5 &  5 & 0    \\
U3755           & 07 13 52 & +10 31 19 & 10 &  5.22    & $-15.36_{M99}$ & 6.9   & 2.6 & 23.4 & 32 & 0    \\
KK65            & 07 42 31 & +16 33 40 & 10 &  4.51 &     $-13.32_{0}$  & 7.1   & 1.2 & 23.3 &  1 & 0    \\
U4115           & 07 57 02 & +14 23 27 & 10 &  5.49 &     $-14.12_{0}$  & 7.4   & 2.9 & 24.8 &  3 & 0    \\
KKH86           & 13 54 34 & +04 14 35 & 10 &  2.61 &     $-11.19_{0}$  & 6.6   & 0.5 & 24.5 &  0 & 0    \\
UA438,E470$-$18   & 23 26 28 & $-$32 23 26 & 10 &  2.23 & $-11.94_{0}$  & 7.4   & 1.0 & 23.2 &  2 & 0    \\
KKH98           & 23 45 34 & +38 43 04 & 10 &  2.45    &  $-11.28_{0}$  & 7.1   & 0.8 & 25.0 &  0 & 0    \\
E321$-$014        & 12 13 50 & $-$38 13 53 & 10 &  3.19 & $-13.18_{0}$  & 6.8   & 1.3 & 24.0 & 0  & 0    \\
\noalign{\smallskip}
\hline
\end{tabular}
\end{table*}

The galaxies were surveyed with the HST Wide Field and Planetary Camera 2
(WFPC2; snapshot programs GO--8192, GO--8601) with 600-second exposures
taken in the F606W and F814W filters for each object. Accurate distances
to 111 nearby galaxies were determined in these snapshot programs
\citep[e.g.][]{kara03} and provide us an unique benchmark to study
properties of globular cluster systems in a number of low surface
brightness dwarf galaxies. \cite{kara01a} searched for globular cluster
candidates in dSph galaxies of the M81 group. We extend this work using
our selection criteria and include also other low-mass galaxies in this
sample. In general, surveying virtually the full area of all our dwarf
galaxies, given their relatively small angular sizes of $\la2$\arcmin,
allows us to study the spatial distribution of globular cluster
candidates.

In the following we briefly discuss a few particularly interesting
galaxies. WFPC2 images with marked GCCs in UGC~3755 and Holmberg~IX are
shown in Figure~1a,b. These two galaxies, which have different properties
and are located in different environments (see Tab.~\ref{tab:prop}), have
the largest numbers of globular cluster candidates among our sample
galaxies. UGC~3755 is an isolated dwarf irregular galaxy \citep{kara04}.
Holmberg~IX is a tidal dwarf companion of M81 \citep{yun94, boyce01}. We
searched for globular cluster candidates in other tidal dwarf companions
of M81, namely BK3N, Arp-loop (A0952+69) and Garland, but found no GCCs in
these galaxies. A faint GCC ($M_{V,0}=-5.2$) in BK3N is located outside
the boundary of the galaxy and probably belongs to M81. Six dSph galaxies
contain GCCs located near their centers. KK84 and KK221 have the largest
number of GCCs among our dSph sample (see Tab.~\ref{tab:prop} and
Figure~\ref{ps:image} c,d). Overplotted on Figure~1d is the isophote of
constant surface brightness $\mu_B\!\sim\!26.5$ mag/arcsec$^2$. All
globular cluster candidates in KK221 are located within the boundaries of
this isophote, but the location of the brightest cluster and the whole
globular cluster system seems to be shifted from the central position in
the galaxy. We can only speculate whether KK221 has an elongated orbit and
experienced strong tidal forces from  NGC5128, which moves its globular
cluster system from a centered position.

Globular cluster candidates were selected using the FIND task of the
DAOPHOT-II \citep{stetson87} package implemented in MIDAS. The detection
threshold was set at 4-$\sigma$ above the background. The minimum full
width at half maximum input parameter (FWHM) used was $\sim\!0.2$\arcsec\
(The stellar FWHM is $\sim\!0.15$\arcsec). Photometry was performed using
PHOT task of DAOPHOT-II with a 12 pixel/1.2\arcsec\ radius. To convert
instrumental magnitudes F606W and F814W into the standard Johnson-Cousins
system we used the surface photometry recipes and equations presented by
\cite{holtzman95}. Surface brightness profiles and growth curves were
computed from the aperture photometry results. The growth curves were
extrapolated to an infinitely large aperture. Finally, the magnitudes of
GCCs were corrected for Galactic extinction using reddening maps from
\cite{schlegel98}, and absolute magnitudes were computed by applying
the distance moduli reviewed in \cite{kara04}. Half-light radii and linear
projected separations of GCCs from the centers of parent galaxies were
converted to a linear measure in parsecs.

We performed photometry of 37 dwarf galaxies using the WFPC2 images and
the surface photometry recipes and equations given in \cite{holtzman95}.
All steps were identical to that used in \cite{mak99} and we refer to
this work for further detail. Integrated absolute $V$ magnitudes are listed
in Table~\ref{tab:prop}.


\section{Cluster Candidate Selection}
\label{ln:selection}
 Our primary target lists include stars, galaxies, and star clusters. In order
to select globular cluster candidates we applied a color selection cut of
$0.3\!<\!(V-I)_0\!<\!1.5$, which is the full range expected for clusters
older than 100 Myr and metallicities $-2.5\!<\!{\rm [Z/Z_{\sun}]}\!<\!0.5$
\citep[e.g.][]{bc03}. We select round objects
(FWHM($x$)$\simeq$FWHM($y$)) with half-light radii of $2\!<$~FWHM~$<\!9$
pix. Reduced to linear measure in parsecs using the distance measurements
of \cite{kara04}, this range corresponds to projected half-light radii
$3\!<\!r_{\rm h}\!<\!20$ pc, which are within values typical for Galactic
globular clusters \citep[][and 2003 update]{harris96}.
Then we performed a visual inspection of selected GCCs on the WFPC2
images. The high angular resolution of HST helps us to reject objects
which show evidence of spiral or disturbed substructure (most likely
background galaxies) from the list of globular cluster candidates.

\begin{figure}[!t]
\includegraphics[width=8.0cm, angle=-90, bb=40 50 540 580]{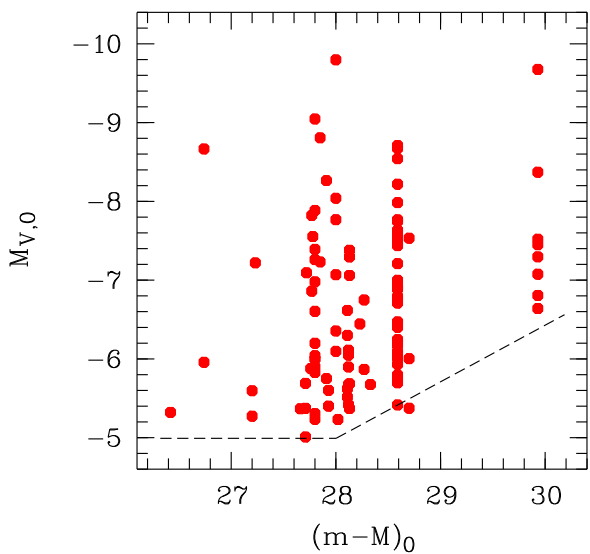}
\centering
\caption{Distance modulus versus absolute $V$ magnitude for GCCs, identified in
this study and listed in Table~\ref{tab:smpl1}. Our sample is complete
down to $M_V\approx-5$ for all our sample galaxies, except KK84.
Detections of GCCs in KK84 are complete down to $M_V \approx-6.6$ mag.}
\label{ps:cmplt}
\end{figure}

We apply a lower absolute magnitude limit for our GCCs of $M_V\!=\!-5.0$
mag. For all our images, this magnitude limit is much brighter than the
photometric limit of the images ($V\!\sim\!25.0$ mag), hence no
completeness corrections are necessary. In other words, our sample is
complete down to $\sim2.5$ mag past the presumed turnover of the
globular cluster luminosity function \citep[e.g.][]{harris01}.
Figure~\ref{ps:cmplt} shows that GCCs detections for the galaxy KK84
located at a distance of 9.7 Mpc is complete down to $ M_{V,0} \approx
-6.6$. Assuming that globular clusters in this galaxy have an intrinsic
luminosity function similar to the Milky Way globular cluster luminosity
function, we detect $\sim 80\%$ of all globular clusters.

Finally, we fit surface brightness profiles to our GCCs with the King law
\citep{king62}
\begin{equation}
\mu_{i} = \left(\frac{1}{\sqrt{1+r^{2}}} - \frac{1}{\sqrt{1+c^{2}}} \right)^{2}.
\end{equation}
We minimize the $\chi^2$ function:
\begin{equation}
\chi^2 = \sum_i\frac{(\mu_i-\overline{\mu}_i)^2}{\sigma^2_i},
\end{equation}
where $\mu_i$ is an average surface brightness inside a circular ring
aperture, $\overline{\mu}_i$ is the predicted value for the same circular
ring, and $\sigma_i$ is the corresponding photometric error. Nonlinear
least-square fits give us the following parameters: $r_c$ --- core radius,
$r_t$ --- truncation radius of the King model, concentration parameter
$c\!=\!r_t/r_c$, and $\mu_0$ --- central surface brightness of the GCC.
The King law approximation provides us
an additional argument to reject background galaxies from our GCCs list.
\cite{king62} emphasized that "relative to globular clusters, giant
elliptical galaxies have an excess of brightness near the center". Hence,
we rejected objects with a central excess brightness and/or uncertain
output parameters. About 10\% of the sample was removed in this way
by visual inspection. Most of the rejected sources have $V-I>1.4$ and are
therefore likely to be background galaxies.

\begin{figure}[!b]
\includegraphics[width=8.0cm,angle=-90, bb=40 50 540 580]{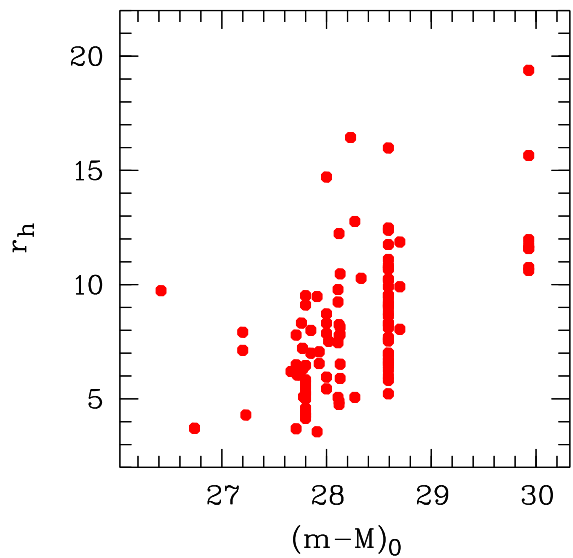}
\caption{Half-light radius of globular cluster candidates (see
Table~\ref{tab:smpl2}) versus distance modulus of their host galaxies. Our
sample is composed mainly of GCCs with core radii $\sim\!3\!-\!13$ pc (see
Section~3 for details).}
\label{ps:obssel}
\end{figure}

The final list of all GCCs in our sample galaxies is presented in
Table~2. Parameters obtained by the King-law approximation
of GCC surface brightness profiles are listed in the
Table~3.

It should be mentioned that, even after applying all our selection
procedures, we can not be certain that all objects in our list are
genuine globular clusters. Elliptical galaxies at intermediate redshifts
are hard to distinguish from globular clusters using $V$ and $I$
magnitudes only. Unresolved starbursts with ages $\le\!300$ Myr at
redshifts $z\!\sim\!0.1\!-\!1.0$ are potential contaminants
\citep{puzia04}. We estimate the number of background galaxies down to
$I=22.5$ mag based on {\it FORS Deep Field} data \citep{heidt03}. We
expect $\sim\!3\!-\!4$ background galaxies within the WFPC2 field of view
with colors resembling those of globular clusters. Eleven of our sample
GCCs are located outside the $\mu_B\!\sim\!26.5$ mag/arcsec$^{2}$
isophotes of their respective host galaxies. The probability that they are
background galaxies is higher than for the other GCCs. Five of these
GCCs are located in dIrr and dSph/dIrr galaxies and have colors $(V-I)_0
>1.2$ and absolute magnitudes within the range  $-5.3<M_{V_0}<-6.3$.
Another six GCCs are located outside the $\mu_B\!\sim\!26.5$
mag/arcsec$^{2}$ isophote, which have $0.6\!<\!(V-I)_0\!<\!1$ and
$-5.2\!<\!M_{V,0}\!<\!-6.6$. Four of these belong to dSphs. Hence, we
estimate the contamination of background objects within the boundaries of
the $\mu_B\!\sim\!26.5$ mag/arcsec$^{2}$ isophote for our galaxies with $
(V-I)_0<1.2$ to be $\la 10\%$.

Before analysing properties of GCCs we investigate our sample for
observational selection effects. Figure~\ref{ps:obssel} shows the
half-light radii of our GCCs (see Table~3) as a function of the distance
modulus of their host galaxy. Our sample is mainly composed of GCCs with
core radii $\sim\!3\!-\!13$ pc. With the current dataset we cannot rule
out the presence of GCCs fainter than $M_{V}\approx-6.5$ and with core
radii $\la10$ pc in the dSph galaxy KK84, which is the most distant host
galaxy in our sample, situated at 9.7 Mpc.

On the other hand, we detect five GCCs with $r_c >12$ pc in galaxies that
are more distant than 3.8 Mpc. This fact might be caused by two reasons.
Firstly, the size of space volume projected into an image pixel increases
with increasing of the distance to galaxies. Errors of structural
parameters grow accordingly. However, we do not find a significant
increase of the mean error for these five objects compared to the rest of
our sample. Secondly, the surveyed area of galaxies grows with distance.
This might lead to a higher detection rate of large GCCs in the outskirts
of these galaxies. Two of these five objects  are located at the largest
galactocentric radii with respect to their host galaxy (see Section 5 for
details). This in turn means that we might miss a few extended globular
cluster candidates in nearby galaxies at large galactocentric radii.
Larger field coverage for nearby systems would help to resolve this issue.
For the remainder of this work, we keep in mind that the lack of extended
GCCs in nearby galaxies is a potential bias of our current dataset.


\section{Properties of Globular Cluster Candidates}
\subsection{Colors}
\label{ln:colors}
\begin{figure}[!t]
\includegraphics[width=8.5cm, bb=0 0 400 400]{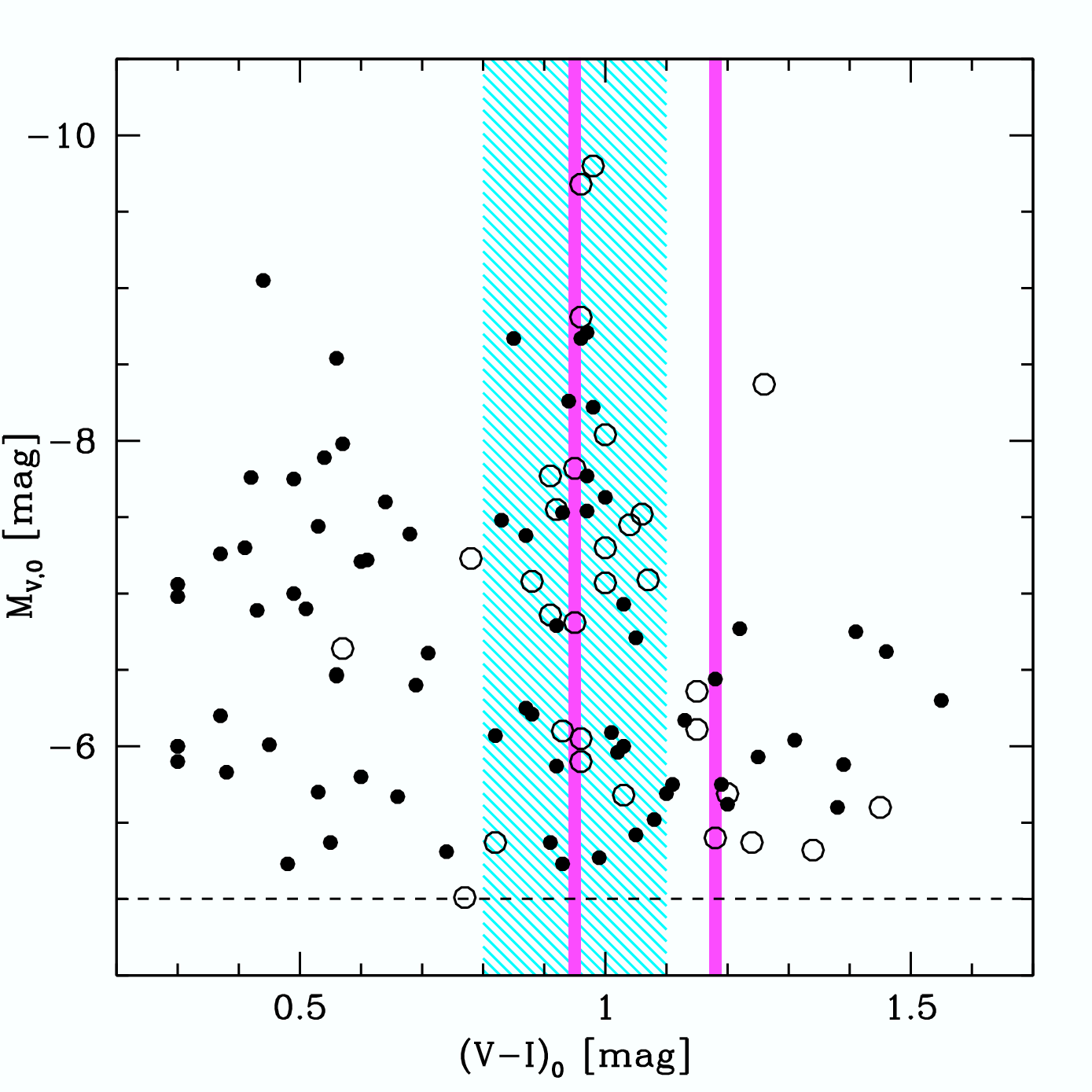}
\centering
\caption{Color-magnitude diagram of globular cluster candidates in dSph ({\it open
circles}) and dIrr galaxies ({\it solid dots}), identified in this study
and listed in Table~2. The dashed line indicates the limit of our photometry
(see Sect.~\ref{ln:selection} for details). Note that there are virtually no GCCs in
dSph galaxies with colors bluer than $(V\!-\!I)_{0}\approx0.7$ mag. The hatched
region shows the color range where most Galactic and M31 globular clusters
are found \citep[see][]{puzia04}. The two vertical lines indicate colors
of blue and red sub-populations in massive early-type galaxies
\citep[e.g.][]{kundu01a, kundu01b, larsen01}.}
\label{ps:cmd}
\end{figure}
\begin{figure}[!t]
\includegraphics[width=8cm, bb=0 0 220 390]{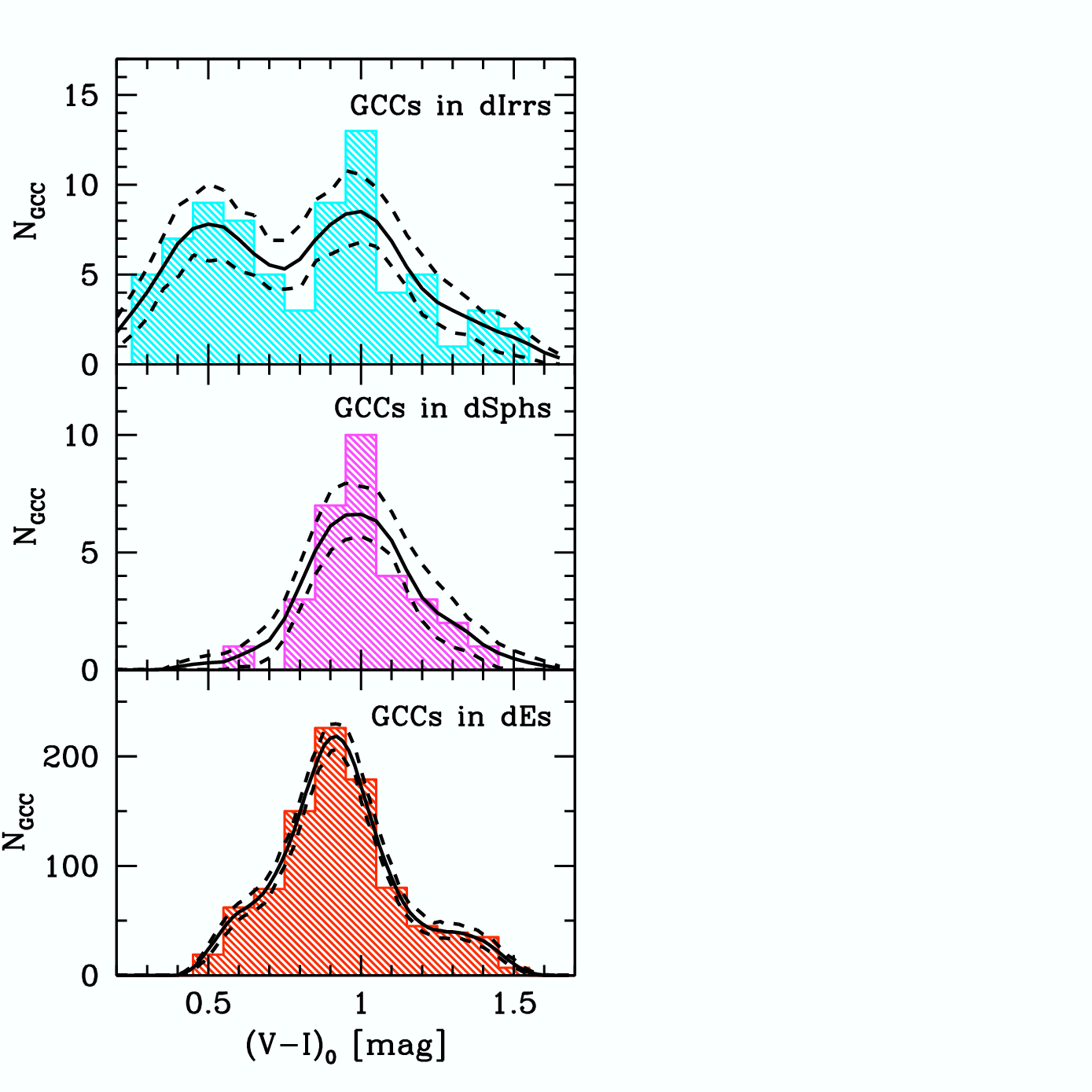}
\centering
\caption{Color distributions of globular cluster candidates in dSph, dIrr, and dE
galaxies. The upper and middle panel show our data, while the bottom
histogram was constructed from data taken from \cite{lotz04}. Solid lines
indicate a non-parametric probability density estimate using an
Epanechnikov kernel \citep{silverman86}. Dashed lines show 90\% confidence
limits.}
\label{ps:vihisto}
\end{figure}

Figure~\ref{ps:cmd} shows a color-magnitude diagram for all GCCs.
Colors and magnitudes were corrected for Galactic foreground extinction
using the reddening maps of \cite{schlegel98}. We have no means to correct
for internal reddening of the observed dIrr galaxies. However, given the
similarity of these systems to nearby dIrr galaxies we estimate that this
correction is $E_{(B-V)}\la0.1$ mag \citep{james05}. Hence, we refer in
the following to foreground extinction corrected magnitudes and colors by
indexing them with a zero. 

A KMM test \citep{ashman94} for GCCs in dIrrs returns peaks at
$(V\!-\!I)_{0}=0.48\pm0.02$ and $1.02\pm0.03$ mag with dispersions $0.12$
and $0.22$ mag for the blue and red peak, respectively. The peak of the
dSph distribution is located at $(V\!-\!I)_{0}=1.01\pm0.03$ mag and has a
dispersion of $0.18$ mag. The red peak GCC sub-population in dIrrs and
most GCCs in dSph galaxies cover the same $(V\!-\!I)_0$ color range as the
ancient Galactic and M31 globular clusters \citep[see Fig.~1
in][]{puzia04}. More than $\sim50$\% of all globular cluster systems in
massive early-type galaxies show indications for multi-modality, with mean
peak colors at $(V\!-\!I)_{0}\approx0.95$ and $\sim1.18$ mag
\citep[e.g.][]{kundu01a, kundu01b, larsen01}. So far most of these
globular cluster systems were found to be old \citep{puzia99, jordan02}. A
comparison of $(V\!-\!I)_{0}$ colors reveals that most GCCs in dSphs and
the red-peak GCCs in dIrrs are similar to {\it blue} globular clusters in
early-type galaxies (see vertical lines in Fig.~\ref{ps:cmd}), which
implies similar ages and metallicities. The blue sub-population of GCCs in
dIrr galaxies is significantly bluer and suggests much younger ages and/or
lower metallicities.

Figure~\ref{ps:vihisto} shows the color histogram for GCCs in dIrr and
dSph galaxies. The distribution of $(V\!-\!I)_{0}$ colors exhibits an
obvious bimodality for GCCs in dIrr galaxies, in contrast to the
distribution of GCCs in dSph galaxies, which shows a single mode
distribution. The clear difference between the two distributions is
underlined by a non-parametric probability density estimate
\citep{silverman86}. The color of the red peak in the dIrr distribution is
virtually identical with the mean of the dSph distribution. Their
dispersions are also very similar. This points to the fact that both
galaxy types host similar globular cluster populations.

We compare the color distributions of GCCs in dIrr and dSphs with that of
dE galaxies. In the bottom panel of Figure~\ref{ps:vihisto} we present the
color histogram of globular cluster candidates in 69 dwarf elliptical
galaxies in the Virgo and Fornax galaxy clusters and the Leo group, with
data taken from \cite{lotz04}. The peak of this distribution is at
$(V-I)_{0}=0.90\pm0.03$ mag. The dE color distribution, in particular the
red end, is similar to the one of dSph GCCs and the red peak of the dIrr
distribution. It is remarkable that the dE color distribution is shifted
to bluer colors by $\sim0.1$ mag with respect to the mean color of the
dSph and the red peak of the dIrr GCC sub-population. This indicates a
mean difference in age and/or metallicity. To better assess the
reality of this color difference, we carried out an independent photometry
of GCCs in one galaxy (VCC1254) from the sample of \cite{lotz04}. For this
purpose we applied our photometric routines to the identical images as
used in the \citeauthor{lotz04} study and compared our results to their work. We
found a small systematic offset $\Delta (V-I)=0.04 \pm 0.05$ for
21 common objects, in the sense that our values tend to be redder. This
indicates that different photometric approaches might have a small
influence on our conclusions. A Kolmogorov-Smirnov test shows that there
is a 0.03\% likelihood that the dE and dSph color distributions were drawn
from the same sample. This likelihood decreases significantly for the
other combinations dE--dIrr and dSph--dIrr.

SSP models \citep[e.g.][]{bc03} predict that for a solar-metallicity, 13
Gyr old stellar population a $\Delta(V-I)_{0}=0.1$ difference translates
into $\sim7$ Gyr younger ages (at the same metallicity) and/or a $\sim0.6$
dex smaller metallicity (at the same age). Because of the age-metallicity
degeneracy of photometric colors, it is difficult to derive individual
ages and metallicities from two colors only. However, a comparison of the
colors with current SSP models shows that $(V\!-\!I)\approx1.0$ mag is
consistent with stellar populations with a certain combination of ages
older than 1 Gyr and metallicities [Z/H]~$\ga-1.4$ dex. The blue peak of
the dIrr color distribution, on the other hand, is consistent with stellar
populations that have ages $\la\!1$ Gyr and metallicities [Z/H]~$\ga-2.0$
dex. Thus, we suggest that red-peak GCCs in dIrr and most GCCs in dSph are
metal-rich globular clusters with intermediate to old ages. Blue GCCs in
dIrr galaxies are likely to be relatively young globular clusters,
probably similar to populous star clusters found in several star-forming
nearby galaxies \citep[e.g.][]{larsen00}.


\subsection{Luminosity Function}

\begin{figure}[!t]
\includegraphics[width=9cm, bb=0 0 400 400]{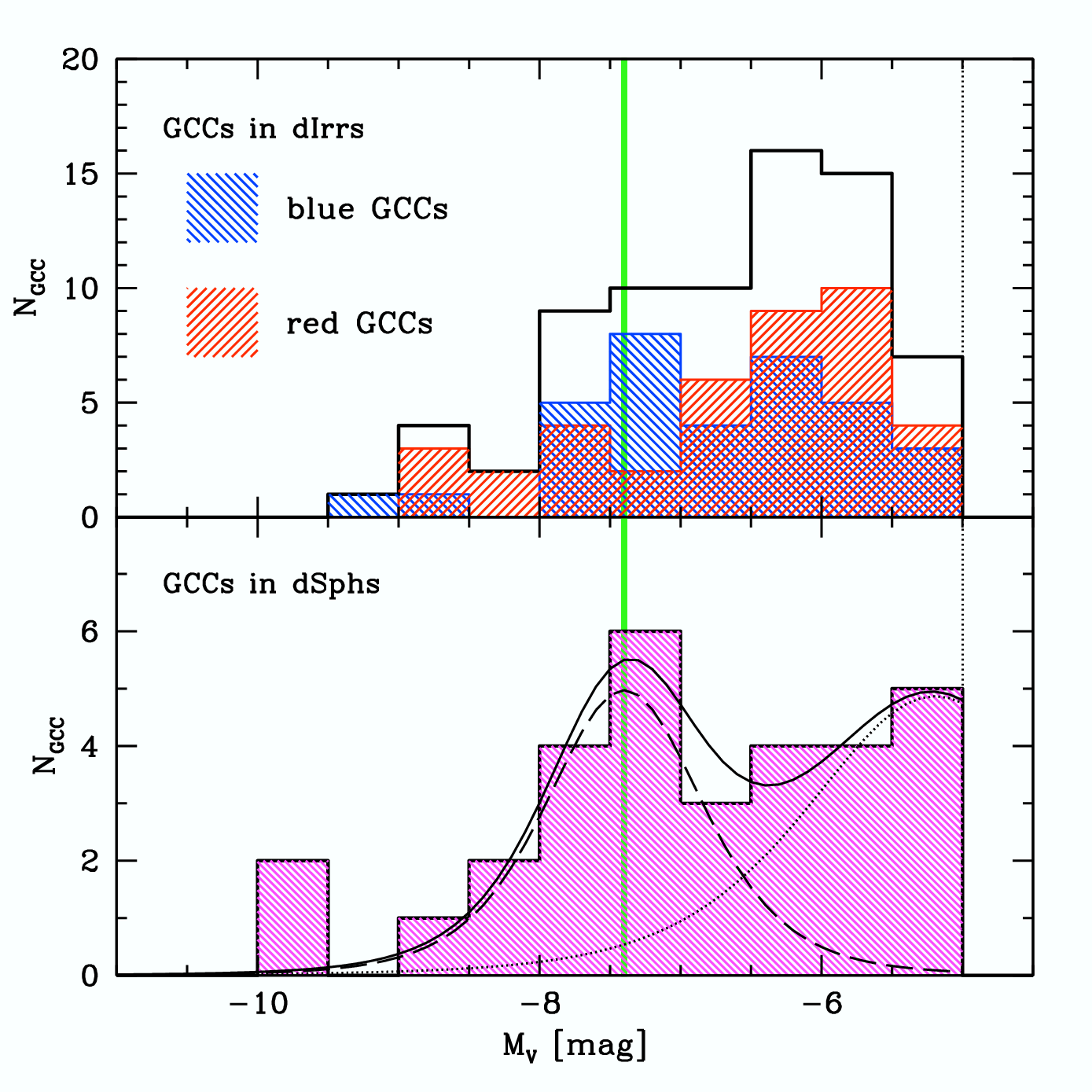}
\centering
\caption{Luminosity functions of globular cluster candidates in dIrr ({\it top
panel}) and dSph galaxies ({\it bottom panel}). The luminosity function of
red and blue GCCs in dIrr galaxies are also shown in as hatched
histograms. The shaded vertical line indicates the turnover of the
Galactic globular cluster luminosity function. The dotted line represents
our photometric limit.}
\label{ps:lf}
\end{figure}

In addition to a blue and likely younger population of GCCs in dIrrs, the
color distributions revealed a red and presumably ancient population in
both dSph and dIrr galaxies. To investigate the luminosity functions (LFs)
of these sub-populations individually, we split the dIrr sample at
$(V\!-\!I)_{0}=0.75$ mag (see Fig.~\ref{ps:vihisto}) into red and blue
cluster candidates. The corresponding LFs are shown in Figure~\ref{ps:lf}
down to $\sim\!2.5$ mag past the turnover of a typical globular cluster
luminosity function, which is indicated by a vertical line \citep{harris01}. 

The LF of blue GCCs seems broad with a turnover somewhere between
$M_{V}\approx-7.5$ and $-6.0$ mag. The fact that we see a turnover for
these supposedly young clusters is remarkable, since constant power-law
slopes down to faint magnitudes are observed in other globular cluster
systems of similar age \citep[e.g.][]{whitmore99, goudfrooij04}. While
these clusters are found in dense environments of ongoing mergers and
massive early-type galaxies, our sample GCCs are located in low-mass
galaxies in loose groups and the field. The reason that we see a turnover
in a young cluster system might be a consequence of fundamentally
different mechanisms of star cluster formation. The star formation rate in
our sample dIrrs is relatively low. Hence, slow spontaneous instabilities
are likely to dominate the star formation process, that do not support the
formation of gravitationally bound low-mass star clusters. Another
possible explanation might be more efficient destruction processes, such
as infant mortality \citep{lada03}, which is thought to be the consequence
of early gas ejection. In general, since physical differences between
young star clusters are related to the pressure differences of the
environment in which they form \citep[e.g.][]{elmegreen97, ashman01},
slight differences in luminosity functions of young globular cluster
systems in different environments cannot be excluded.

The LF of red GCCs in dIrrs turns over at a much fainter magnitude
$M_{V}\approx-6$ mag than for their blue counterparts. We can only
speculate whether this is due to a difference in formation mechanisms,
ages/metallicities, or simply due to contamination by background galaxies.
We point out that contamination by stellar crowding is more likely to
occur in dIrr galaxies, which are populated by bright stars. However,
stellar crowding is unlikely to explain the excess of objects, which is
also found in dSph galaxies (see below).

For GCCs in dSphs the luminosity function shows a turnover at around
$M_V=-7.4$ mag. Similar turnover magnitudes are also found in many other
globular cluster systems \citep{harris01}. In addition to a roughly
log-normal LF for bright objects, we find that the luminosity functions of
GCCs fainter than $M_V=-6.5$ mag are steadily increasing towards fainter
magnitudes, possibly resembling a dynamically less evolved population of
clusters \citep{gnedin97, fall01}. The existence of an excess population
of faint globular cluster candidates in dSphs is surprising and is
reported on here for the first time. A scaled {\it Student} $t_5$-distribution 
\begin{equation}
t_{5}(m|M_{V,0},\sigma_t)=\frac{8}{3\pi\sigma_{t}\sqrt{5}}\left(1+\frac{(m-M_{
V,0})^2}{5\sigma_t^2}\right)^{-3}, 
\end{equation} 
that is the best approximation to the observed Galactic globular cluster
luminosity function \citep[][see dashed line in the bottom panel of
Fig.~\ref{ps:lf}]{secker92}, shows good agreement with the bright
peak\footnote{This approximation neglects two very bright GCCs with
magnitudes $M_{V}\approx-10$.}, where $M_{V,0}=-7.4$ and $\sigma_{t}=0.6$.
The faint cluster excess can be well approximated by a second $t_5$-peak
with a mean at $M_{V,0}\approx-5.2$ mag and roughly twice as broad a
dispersion as the bright peak. This is reminiscent of the composite
luminosity function of globular and open clusters in the Milky Way.
However, since this is a composite luminosity function for GCCs in
different galaxies, various dynamical processes are lumped together in one
sample. This and the possibility of variable internal reddening might
account for at least part of the spread in the observed luminosity
function.

We note that a similar excess population of faint star clusters has been
discovered in the low-mass spiral M33 \citep{chandar01} and in nearby
lenticular galaxies \citep{brodie02}. However, the nature of these faint
excess clusters is difficult to assess based on two-color photometry.
Bright nuclei of barely resolved background galaxies might contribute to
the faint GCC excess.

For KK84 and UGC~3755, which host the richest GCC systems in their
morphology class, we investigate their GCC luminosity distributions
individually. The GCC luminosity function of the dSph galaxy KK84 peaks
near $M_{V} \approx-7.3$ mag. For the dIrr galaxy UGC~3755 we find peaks
near $M_{V}\approx-7.0$ mag for both the blue and red sub-population. Both
galaxies are relatively massive compared to the rest of the sample and
show typical turnover magnitudes.

In summary, given the number statistics of our data, we can say that,
compared to a typical globular cluster luminosity function, both the dSph
and dIrr GCC populations exhibit an excess of faint globular cluster
candidates. Whether this is due to a genuine new population of low-mass
star clusters or due to contamination by marginally resolved background
galaxies will be resolved with spectroscopic data.

\begin{figure}[!t]
\includegraphics[width=5.9cm,angle=-90,bb = 60 60 540 720]{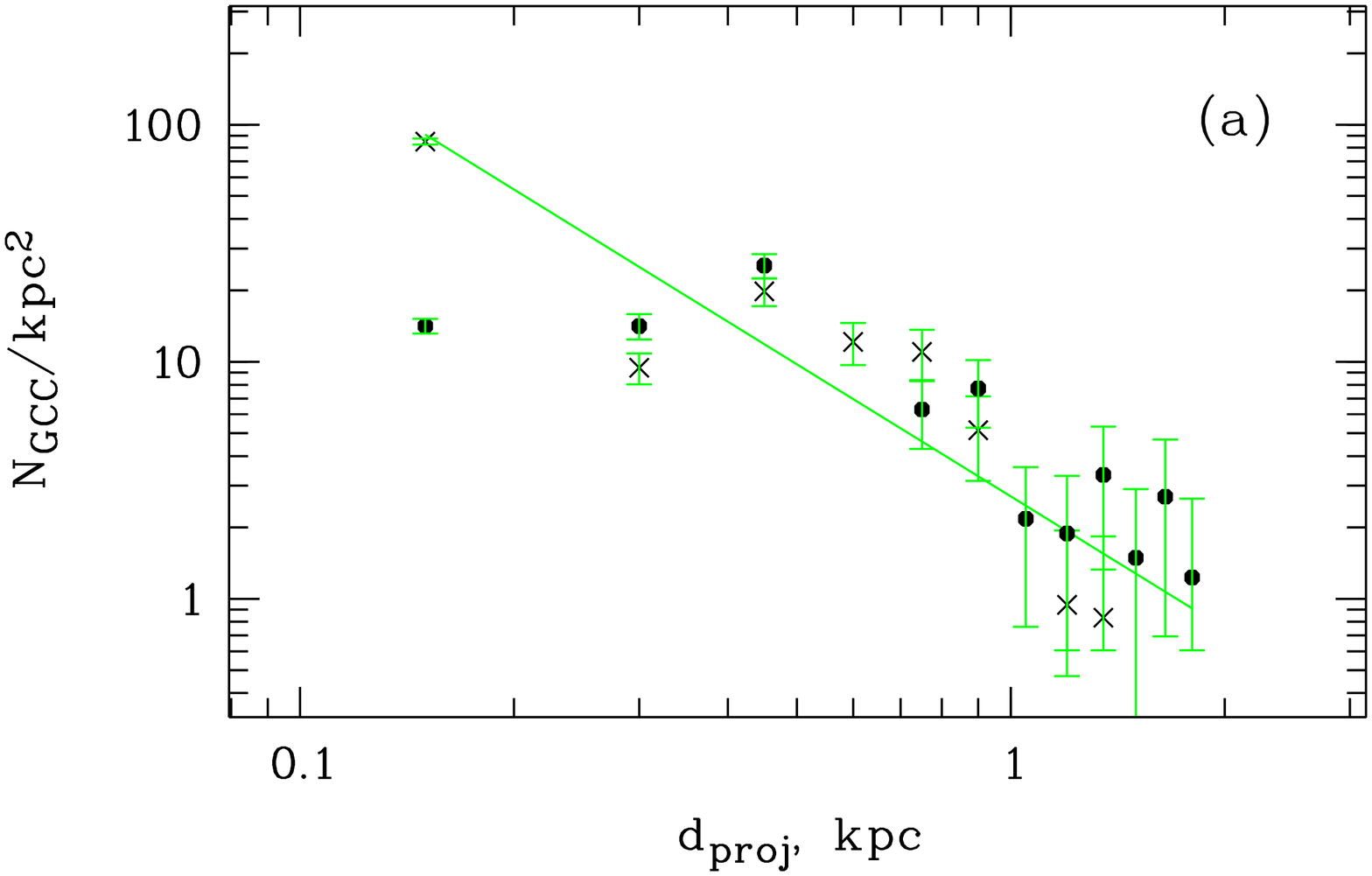}
\includegraphics[width=5.9cm,angle=-90,bb = 60 60 540 720]{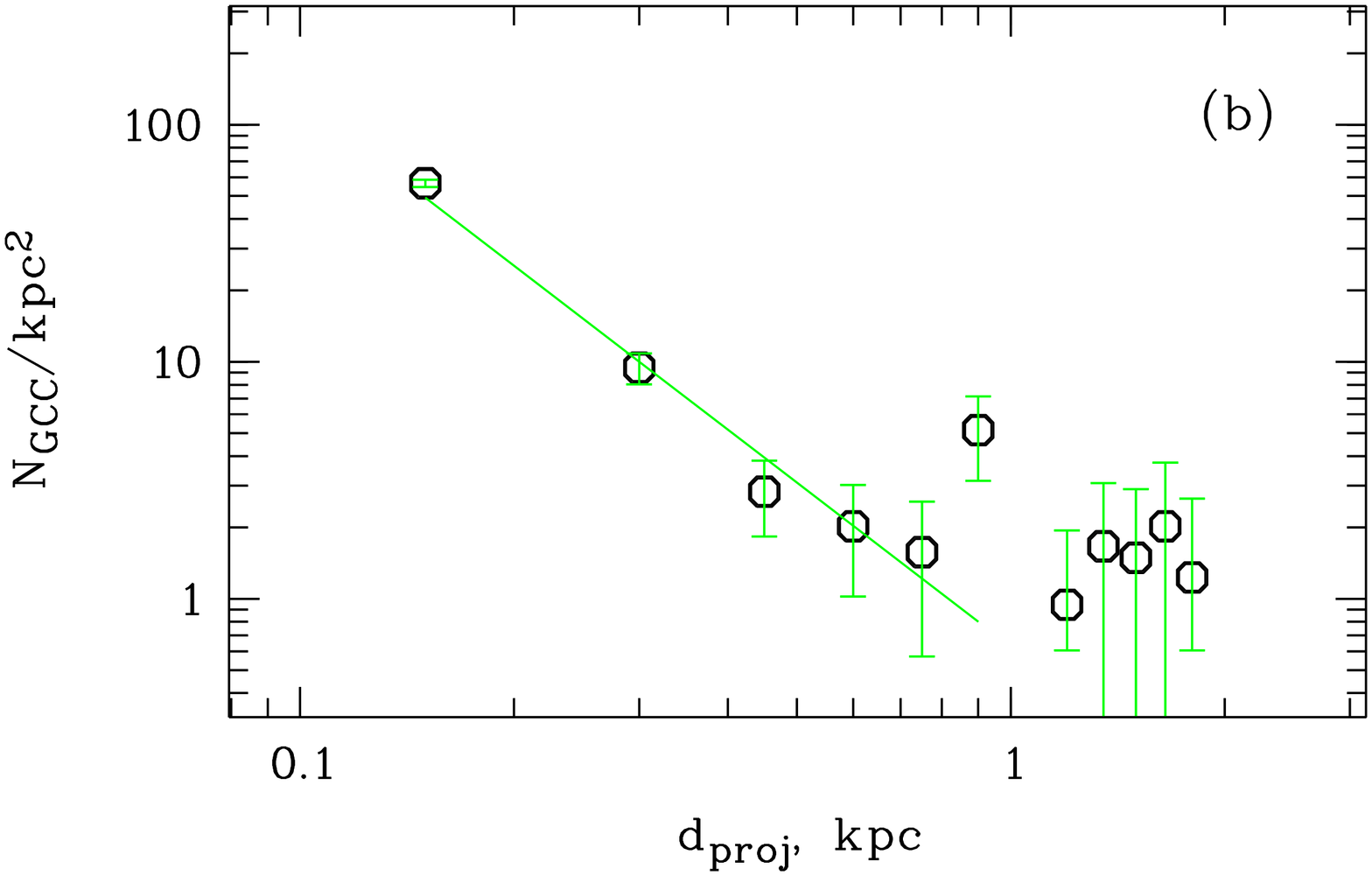}
\caption{Radial distribution of GCCs in dIrr galaxies {\it (panel a)} and dSph
galaxies {\it (panel b)}. The sample of GCCs in dIrrs is divided into blue
GCCs with $(V-I)_0<0.75$ {\it (dots)} and red GCCs $ (V-I)_0>0.75$ {\it
(crosses)}. Plotted here is the logarithm of the surface density of GCCs
per square kpc (evaluated in 0.15-kpc bins) vs. the logarithmic projected
distance from the galaxy center, in kpc.}
\label{ps:rad_dist}
\end{figure}


\subsection{Spatial Distributions}
\label{ln:spatial}
In the following, we compare the composite spatial distribution of GCCs in
dSph and dIrr galaxies. We divide the population of GCCs in dIrr galaxies
into red and blue objects at $(V-I)_{0}=0.75$ mag. In
Figure~\ref{ps:rad_dist} we plot the logarithmic number density of GCCs
per square kpc versus the logarithm of the linear projected separation
from the galaxy center in kpc, for dIrr and dSph galaxies. Radial
distributions of blue and red GCCs in dIrrs are shown by different symbols
and reveal somewhat different slopes. A power law for the surface density
of the form $\rho= r^{-x}$ gives a good fit to our data. We find that the
surface density profiles of globular cluster systems in dIrr galaxies are
"flatter" ($x_{\rm red} \approx 1.1$) and ($x_{\rm blue} \approx 1.85$)
than those in dSph galaxies ($x \approx 2$). The profile for GCCs in dSph
galaxies seems to flatten out beyond $\sim\!1$ kpc galactocentric
distance. Hence, the fit includes only the inner part of the GCC
population. The profiles of the GCC population in dIrr galaxies do not
show a flattening at large galactocentric distances.

These values are in good agreement with those found for globular cluster
systems in elliptical \citep[$x=0.8-2.6$,][]{puzia04} and dwarf elliptical
galaxies \citep[$x=1.6\pm0.4$,][]{durrell96}. Moreover, \cite{harris86}
found $x=3.5\pm0.5$ for the combined spatial distribution of globular
clusters in NGC~147, NGC~185 and NGC~205. \cite{minniti96} constructed a
composite dwarf elliptical galaxy from all early-type dwarf galaxies in
the Local Group and computed $x=2.1$.


\subsection{Structural Parameters}
\label{ln:structpars}
\begin{figure*}[!t]
\centering
\includegraphics[width=6.1cm,angle=-90]{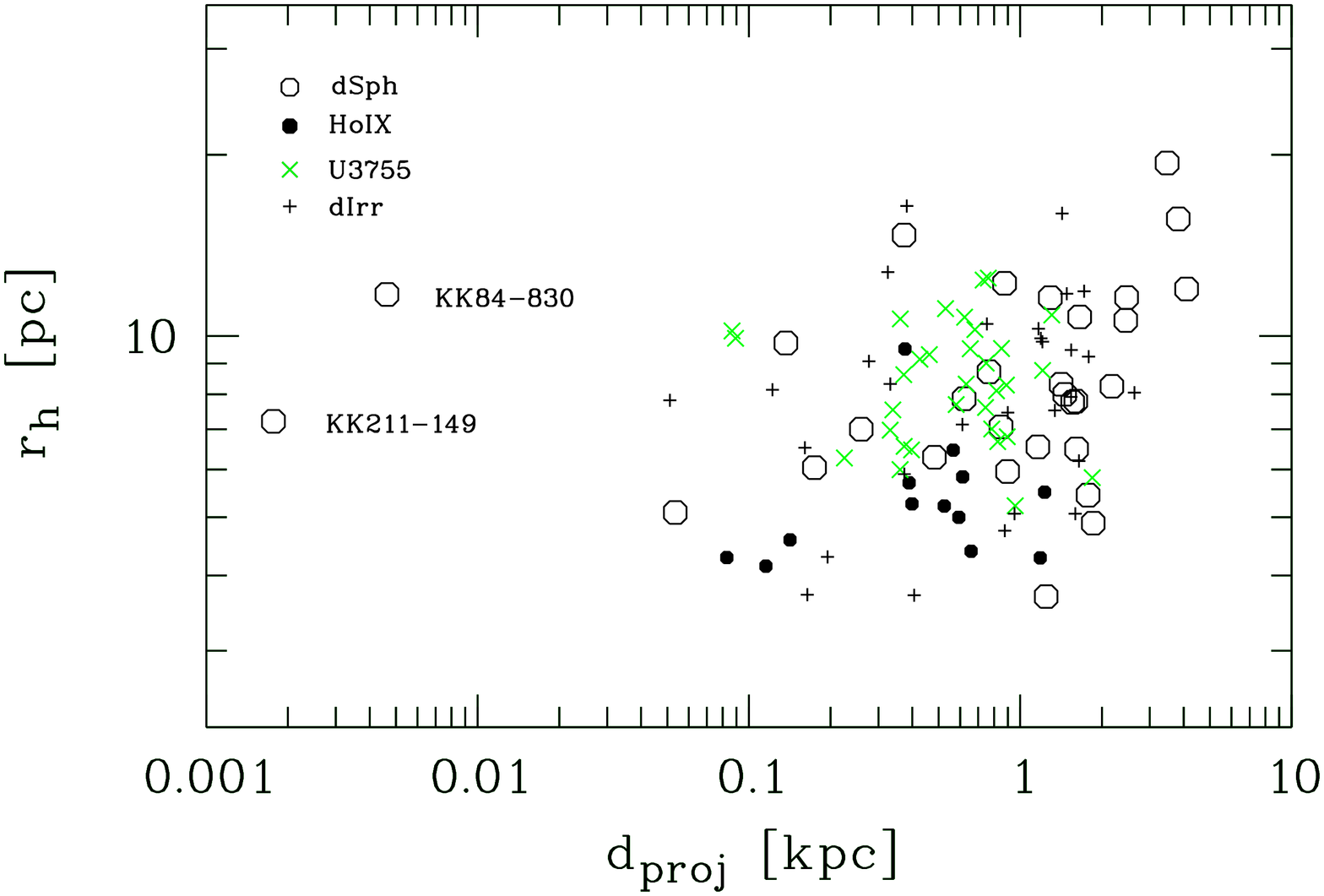}
\includegraphics[width=6.1cm,angle=-90]{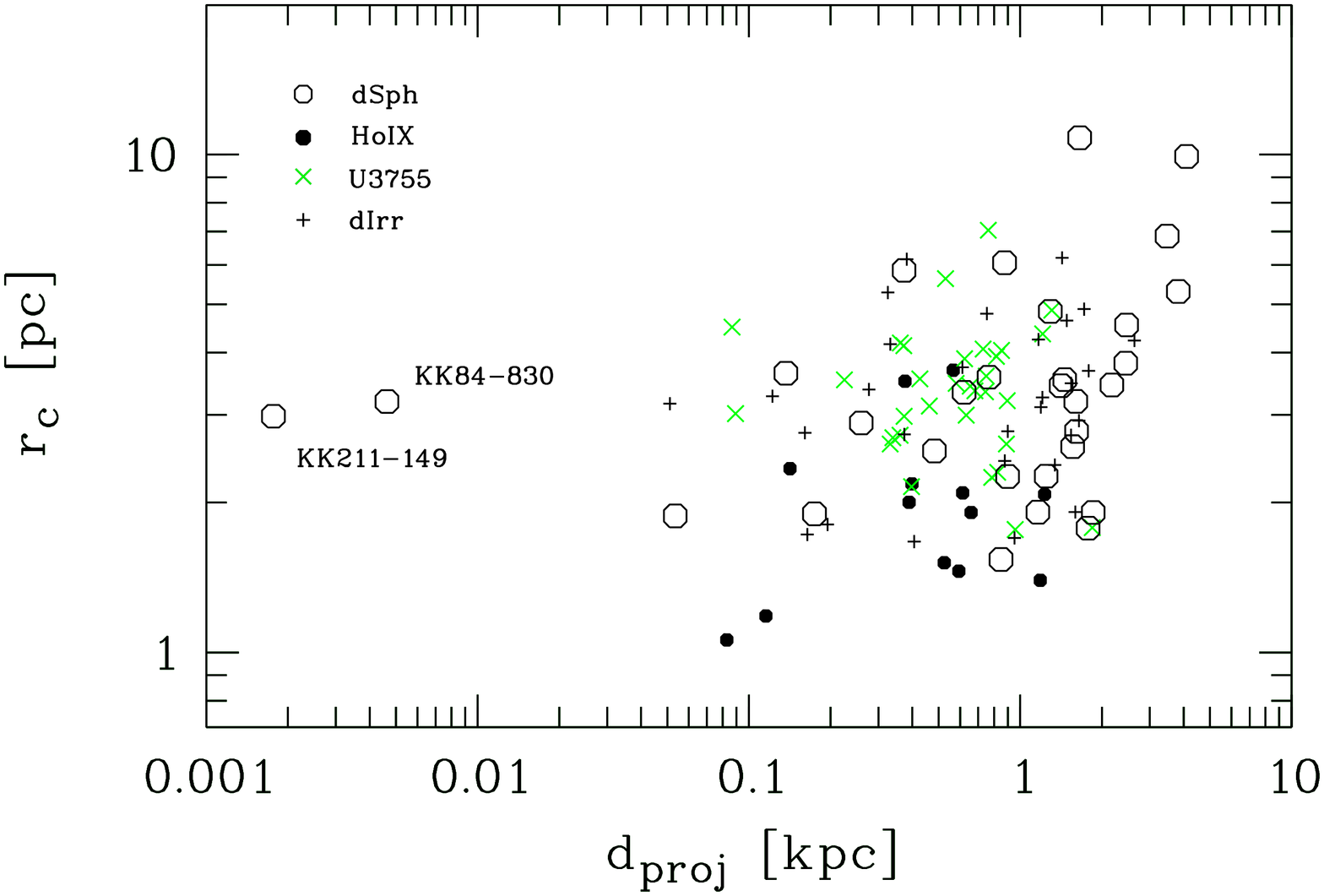}
\caption{{\it Left panel:} Half-light radius of our sample GCCs versus their
projected galactocentric distance. {\it Right panel:} Core radius of GCCs
versus their projected galactocentric distances. The data for the two
richest GCC systems, the dIrr galaxies UGC~3755 and 
Holmberg~IX, are marked by different symbols.}
\label{ps:rade_dep}
\end{figure*}

Under the superb spatial resolution of HST/WFPC2, typical globular
clusters \citep{harris96} begin to be resolved for galaxies less distant
than $D\!\sim\!10$ Mpc. Using our King-profile approximation routine we
measure structural parameters for all our sample GCCs.
Figure~\ref{ps:rade_dep} shows half-light radii, $r_{h}$, and core radii,
$r_{c}$ of GCCs as a function of their projected galactocentric distance.
Both panels show a trend of increasing half-light and core radius as a
function of increasing galactocentric distance. These correlations are
however driven by the outermost GCCs. Spectroscopy is necessary to measure
their radial velocities and test whether these objects are genuine
globular clusters or resolved background galaxies. If we consider GCCs
with projected distances less than $\sim\!1$ kpc to avoid potential
contamination by background sources (see Fig.~\ref{ps:rad_dist}b), we find
only tentative evidence for a $r_{c}\!-\!d_{\rm proj.}$ correlation. With
the same radial constraint we find no correlation between half-light
radius and galactocentric distance.

Such correlations exist for half-light radii and core radii of Galactic
and Large Magellanic Cloud (LMC) globular clusters \citep{vdB00, grijs02,
vdB04}. We find that at a given galactocentric distance our sample GCCs
have on average a factor of $\sim\!5$ larger half-light radii than LMC
globular clusters. This might be due to the higher mass of LMC, which has
a stronger tidal field in which destruction processes are enhanced
compared to our sample dIrr galaxies. The current dataset reveals no
difference between the average structural parameter distribution of GCCs
in dIrr and dSph galaxies.

We detect two GCCs in dSph galaxies at small galactocentric radii with
structural parameters significantly larger than what one would expect from
the extrapolation of the remaining sample. These clusters might be fainter
analogues of nuclear star clusters found in dwarf elliptical galaxies
\citep{durrell96}, that spiraled into the cores of their host galaxies
through the process of dynamical friction and orbital decay
\citep{lotz01}. These two GCCs are among the brightest objects in our
sample ($M_{V}=-9.7$ and $-7.8$ mag, see Tab.~\ref{tab:smpl1}), but they
are significantly larger and fainter than nuclear star clusters in
late-type spiral galaxies, which have typical sizes between
$r_{h}\approx2$ and 10 pc and magnitudes between $M_{V}\approx-10$ and
$-13$ mag \citep{boeker04}. These two nuclear GCCs are also much smaller
than the cores of ultra-compact dwarf galaxies, which were recently
discovered in the Fornax galaxy cluster \citep{hilker99, drinkwater03}.

\begin{figure}[!t]
\centering
\includegraphics[width=8.5cm,angle=-90]{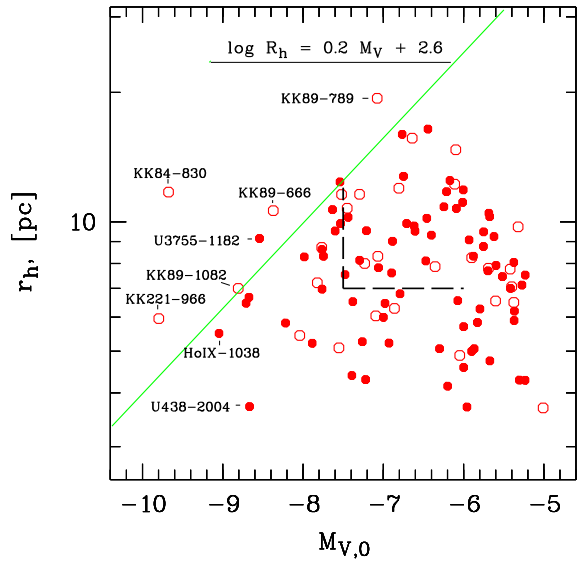}
\caption{Half-light radii of GCCs in dSph ({\it open circles}) and dIrr galaxies
({\it solid dots}) versus their luminosities, $M_{V,0}$. The figure shows that
almost all GCCs lie below the relation for Galactic globular clusters
\citep[{\it solid line}, see][]{vdB04}, except some GCC in dSphs and 
one GCC in UGC3755. The most
interesting cases of bright and compact GCCs in dIrrs and bright and
diffuse GCCs in dSphs are labeled. Roughly half of GCCs is
consistent with luminosities and strucrural parameters of the ``faint
fuzzies'' of \cite{brodie02}, whose location is indicated by two dashed
lines ($M_V>-7.5$ and $r_h>7$ pc).}
\label{ps:Mv_Rh0}
\end{figure}

We also consider structural parameters of the newly discovered population
of faint and extended star cluster in lenticular galaxies, termed ''faint
fuzzies'' \citep{brodie02}, which have typical half-light radii $r_{h}>7$
pc and magnitudes fainter than $M_{V}=-7.5$ (see Figure~\ref{ps:Mv_Rh0}). 
In fact we find that roughly
half of our sample is consistent with their magnitudes and structural
parameters. The colors of these ''faint fuzzy'' clusters are around
$(V-I)_{0}\approx1.3$. We find several faint and extended GCCs in our
sample with very similar colors. However, the majority is bluer and has
colors typical of the red sub-population with a mean at
$(V-I)_{0}\approx1.0$ mag. We note that the most extended globular
clusters in the Local Group spirals and the Magellanic Clouds resemble
these faint fuzzies as well.

Figure~\ref{ps:Mv_Rh0} shows the distribution of our GCCs in the
half-light radius versus luminosity plane. A large fraction of GCCs
exhibits luminosities and half-light radii consistent with ``faint fuzzy''
star clusters. The majority of GCCs, however, fall below the relation
found by \cite{vdB04}, where almost all of the Galactic globular clusters
reside. Moreover, the figure shows that some GCCs, primarily those in dSph
galaxies, show luminosities and structural parameters similar to those
found for NGC~2419 and $\omega$Cen, two atypically extended Galactic halo
globular clusters \citep{vdB04}.

\subsection{Compact Star Clusters in Holmberg IX?}

We find that GCCs in the dIrr galaxy Holmberg~IX are on average more
compact than those situated in dSph and other dIrr galaxies (see
Fig.~\ref{ps:rade_dep}). Almost all Holmberg~IX GCCs are blue with a mean
$(V-I)_{0}=0.45\pm0.16$ and are likely to be young ($t\la1$ Gyr). The
youth of these clusters could be the reason for their compactness, as
either cluster destruction processes did not have enough time to
significantly alter their sizes and/or the initial conditions during
cluster formation in Holmberg~IX were different from those in other
galaxies, for instance due to significantly higher ambient pressure.

In a study of structural parameters of LMC globular clusters,
\cite{grijs02} found that older clusters exhibit a much greater spread in
core radii than their younger counterparts. If the increased spread is due
to advanced dynamical evolution, one would expect a smaller spread in
sizes of old globular clusters in less massive galaxies. We will test this
hypothesis once accurate spectroscopic ages and chemical compositions
become available.


\section{Discussion}
\subsection{Extended Globular Clusters}
There is increasing evidence in the current literature that low-mass
galaxies host a significant population of faint and extended globular
clusters. M33 is the most prominent example \citep{chandar01, chandar04}.
Different physical mechanisms tend to destroy clusters with time:
evaporation, disk shocking, tidal shock heating, dynamical friction etc.
\citep[e.g.][]{gnedin97, surdin98, fall01}. The efficiency of tidal
disruption increases significantly with the presence of a bulge component
\citep{gnedin97}. M33 as well as the dwarf galaxies in our sample have no
significant bulge component and faint and extended globular clusters are
able to survive significantly longer than in the Milky Way or M31.
Consequently, there are no indications for an excess population of faint
and extended globular clusters in the Milky Way or the inner halo of M31
\citep{harris01, barmby01, vdB04}, as the efficiency of cluster disruption
steeply drops with galactocentric distance.

If accretion of globular clusters plays an important role in the assembly
of outer globular cluster systems in massive galaxies, one should expect
similarities in magnitudes and structural parameters between clusters in
our sample dwarf galaxies and the outskirts of the Local Group spirals.
Indeed, most of our GCCs have similar structural parameters and
luminosities as the most extended Galactic and M31 globular clusters,
which are located at large galactocentric distances. This hints at the
accretion of such extended clusters from satellite dwarf galaxies
\citep[see also][]{forbes04}. However, most of the outer halo Galactic
globular clusters are metal-poor ([Z/H]~$\la-1$) and old ($t\ga10$ Gyr).
If the accretion scenario for the outer-halo Galactic globular clusters is
correct, then they must have formed early, perhaps in Searle-Zinn-type
proto-galactic clumps \citep[see][]{searle78}.

It was recently found that the chemical composition of LMC globular
clusters is not entirely reflected in the globular cluster systems of
Local Group spirals \citep[e.g.][]{puzia04b}. Hence, the scenario of the
assembly of their outer-halo globular cluster systems by accretion of
LMC-type globular clusters seems less likely. A detailed spectroscopic
investigation of globular clusters in dwarf galaxies will certainly help
to constrain the picture of globular cluster accretion from satellite
dwarf galaxies, and will also provide insight into the assembly history
of globular cluster systems in giant elliptical galaxies.


\subsection{Star Clusters in Dwarf Galaxies}
Our sample consists of the lowest-mass nearby dwarf galaxies and is
representative and homogeneous enough to study the presence of GCCs as a
function of general galaxy properties, such as morphological type. The
faintest galaxies in our sample are dSphs, with mean surface brightnesses
as faint as $\mu_{\rm B}\!\approx\!24.4$ mag/arcsec$^2$ (see
Tab.~\ref{tab:prop}). Roughly half of our dIrr galaxies are in the same
magnitude range. The detection rate of GCCs in these faint dIrr is
$\sim\!2$ times lower than in dSphs (and the other higher surface
brightness dIrrs). Only 5 out of 19 faint dIrr galaxies contain GCCs,
whereas 10 of 18 dSphs harbor GCCs, at similar host galaxy luminosities.
It is not known what causes this discrepancy.  We speculate that dSph
galaxies turned their gas reservoir earlier and more efficiently into
stars than dIrr galaxies of similar luminosity. This was perhaps due to a
higher virial density (from stars, gas and dark matter) and higher ambient
pressure in the early environments of dSph galaxies. Another reason might
be the different stellar $M/L$ ratio in dSph and dIrr.

It should be mentioned, that our sample dSphs reveal an outstanding
feature: in contrast to dIrr galaxies, $\sim 60\%$ of dSphs with GCCs have
cluster candidates located near their galaxy center. These GCCs are bright
and compact ($-7<M_{V,0}<-10$, $r_c \approx 2-3$ pc), similar to those
found at the center of nucleated Virgo dEs \citep{lotz04}.

It was shown in Section~\ref{ln:spatial}, that the surface density
profiles of globular cluster systems in dSph galaxies are steeper than
those in dIrrs. In general, we find that GCC systems of dSph galaxies are
more spatially concentrated than in dIrr galaxies. The red sub-population
of GCCs in dIrrs shows the "flattest" profile. Blue GCCs in dIrr galaxies,
on the other hand, have the tendency to reside in central regions of their
host galaxies, but they are still less concentrated than GCCs in dSphs.
This difference implies that globular cluster formation and/or evolution
histories in both galaxy types were spatially not alike. Globular clusters
presumably form where gas is undergoing agitated star formation. With
respect to star formation histories, both dSphs and dIrrs appear to be
inhomogeneous classes of galaxies \citep[e.g.][]{mateo98}. The difference
in GCC surface density profiles might reflect a bias of powerful star
formation events towards the center of dSph and/or more efficient orbital
decay of star clusters in these galaxies \citep[e.g.][]{lotz01}.

Both the dIrr and dSph galaxies share a sub-population of red globular
clusters with very similar mean $(V-I)_{0}$ colors. In addition to this
GCC population, dIrr galaxies host a very blue population of clusters,
that are likely to be younger. The bimodality of the GCC color
distribution in dIrr galaxies implies two different episodes and/or
mechanisms of cluster formation. Similar GCC colors as in dSphs are found
for GCCs in dE galaxies (see Sect.~\ref{ln:colors}). There is however a
very puzzling offset of $\Delta(V-I)_{0}\approx0.1$ between GCCs in dE and
dSph galaxies, where GCCs in dE are bluer. This can be interpreted as the
result of lower metallicities and/or younger ages.

The question for the reason of this age and/or metallicity difference is
difficult to answer. However, one can consider the globular cluster system
of the Fornax dSph galaxy in the Local group as a first reference to learn
more about ages and metallicities of field and cluster stellar
populations. Red giant branches of all globular clusters in Fornax dSph
show steeper slopes than the mean RGB slope of the field stellar
population \citep[e.g.][]{buonanno98, buonanno99}. Using RGBs of Galactic
globular clusters as reference, all globular clusters in Fornax dSph were
found to have low metallicities. This was confirmed by a spectroscopic
study of their integrated-light \citep{strader03}, which found a mean
metallicity of [Fe/H]~$\approx-1.8$ and old ages. First attempts to obtain
spectroscopic age and metallicity estimates for globular clusters in dEs
are on the way \citep{miller04}.

From a study of surface brightness fluctuations, \cite{jerjen04} showed
that Virgo dE galaxies follow a galaxy metallicity--luminosity relation.
Virgo dEs should therefore be expected to have on average $\sim\!0.7$ dex
higher metallicities compared to the 100 times fainter LG dSph galaxies.
Assuming old ages for their GCCs in dEs, \cite{lotz04} find a correlation
between the mean globular cluster color and host galaxy luminosity, which
implies a globular cluster metallicity-galaxy luminosity relation of the
form $\langle Z_{\rm GCC}\rangle\propto L^{0.22\pm0.05}_{B}$. This
relation implies bluer globular clusters colors in fainter galaxies.
Globular clusters have generally bluer colors than their host galaxies
\citep{az98}, and the two relations suggest that this offset is present at
all galaxy luminosities. According to those relations, globular clusters
in dEs are expected to have higher metallicities than in dSph galaxies.
Therefore, {\it younger ages} might be responsible for the bluer colors of
globular clusters in dE galaxies.

It is important to continue the study of globular cluster systems in
nearby low-mass galaxies and to compare their properties as a function of
different host galaxy properties, e.g. morphological type and
environmental density.

\acknowledgements
MES and DIM are partially supported by a RFBR~04-02-16115 grant. THP is
supported by an ESA Research Fellowship, which is gratefully acknowledged.
DIM thanks the Russian Science Support Foundation. MES kindly thanks
I.D.~Karachentsev for initiating the beginning of a work on searches of
globular clusters in dSph galaxies and Vladimir Surdin for useful
discussions. We thank Jennifer Lotz for providing data of globular cluster
candidates in dwarf elliptical galaxies in electronic form. THP thanks
Rupali Chandar, Nicole Homeier, and Jennifer Lotz for useful discussions.
We thank the anonymous referee for helpful comments.

\begin{table*}
\centering
\caption{List of globular cluster candidates (GCCs) in nearby LSB dwarf galaxies.
The columns contain the following data: Identifier of each cluster,
composed of the (name of its host galaxy)-(WFPC2 chip)-(cluster
numbering), X, Y coordinates derived from the WFPC2 frames, equatorial
coordinates (J2000.0), half-light radius $r_{h}$ in parsecs, apparent
axial ratio $e=1 - b/a$, integrated absolute $V$ magnitude (corrected for
Galactic extinction using \cite{schlegel98} maps) and corresponding error,
integrated absolute $V-I$ color (corrected for Galactic extinction) and
corresponding error, and the projected separation from the center of its
host galaxy $d_{\rm proj}$ in kiloparsecs. The numbers of GCCs located
outside the isophote of constant surface brightness $\mu_B\!\sim\!26.5$
mag/arcsec$^2$ of the host galaxy are marked by an asterisk symbol ($^*$).}
\scriptsize
\label{tab:smpl1}
\begin{tabular}{lcclrccl} \\ \hline\hline\noalign{\smallskip}
GCC             &     X, Y           &    RA (J2000) DEC     & $r_h$&  $e$  &  $V_{0}$           & $(V-I)_0$    &   $d_{\rm proj}$  \\
\noalign{\smallskip}\hline\noalign{\smallskip}
DDO53$-$3$-$1120    &   291.022, 208.595 &  08 34 04.1 +66 10 23&  6.7 &  0.2 & $-$5.88 $ \pm$ 0.07 &1.39 $  \pm$  0.10&   0.33  \\
BK3N$-$2$-$863$^*$  &   546.324, 579.284 &  09 54 00.1 +68 58 54&  6.8 &  0.1 & $-$5.23 $ \pm$ 0.08 &0.93 $  \pm$  0.11&   1.34  \\
KDG73$-$2$-$378$^*$ &   377.384,  64.881 &  10 53 07.7 +69 32 02&  8.3 &  0.1 & $-$5.75 $ \pm$ 0.08 &1.11 $  \pm$  0.11&   1.54  \\
KK77$-$4$-$939      &    97.291, 330.684 &  09 50 00.3 +67 31 10&  3.7 &  0.3 & $-$5.01 $ \pm$ 0.09 &0.77 $  \pm$  0.12&   1.25  \\
KK77$-$4$-$1162     &   138.177, 568.545 &  09 49 56.2 +67 31 11&  6.5 &  0.1 & $-$5.37 $ \pm$ 0.08 &0.82 $  \pm$  0.11&   1.61  \\
KK77$-$4$-$1165     &   491.903, 572.199 &  09 49 54.8 +67 30 37&  7.8 &  0.0 & $-$5.69 $ \pm$ 0.08 &1.20 $  \pm$  0.11&   1.60  \\
KDG61$-$3$-$1325    &   363.770, 403.865 &  09 57 02.8 +68 35 35&  4.7 &  0.1 & $-$7.55 $ \pm$ 0.07 &0.92 $  \pm$  0.10&   0.05  \\
KDG63$-$3$-$1168    &   347.918, 329.025 &  10 05 07.2 +66 33 30&  6.0 &  0.0 & $-$7.09 $ \pm$ 0.07 &1.07 $  \pm$  0.10&   0.17  \\
DDO78$-$1$-$167     &   421.025, 524.751 &  10 26 28.3 +67 40 45&  7.4 &  0.1 & $-$7.23 $ \pm$ 0.07 &0.78 $  \pm$  0.10&   1.46  \\
DDO78$-$3$-$1082    &   483.889, 549.646 &  10 26 27.1 +67 39 10&  7.0 &  0.1 & $-$8.81 $ \pm$ 0.07 &0.96 $  \pm$  0.10&   0.26  \\
BK6N$-$2$-$524      &   225.256, 182.777 &  10 34 29.3 +66 01 29&  4.4 &  0.3 & $-$5.40 $ \pm$ 0.08 &1.18 $  \pm$  0.11&   0.85  \\
BK6N$-$4$-$789      &   608.148, 247.233 &  10 34 19.8 +66 00 32&  4.5 &  0.3 & $-$5.60 $ \pm$ 0.07 &1.45 $  \pm$  0.10&   1.16  \\
Garland$-$1$-$728   &   477.284, 107.876 &  10 03 30.8 +68 41 55&  3.2 &  0.1 & $-$8.26 $ \pm$ 0.07 &0.94 $  \pm$  0.10&   1.23  \\
HoIX$-$3$-$866      &   384.173, 122.948 &  09 57 37.7 +69 02 29&  5.2 &  0.0 & $-$7.89 $ \pm$ 0.07 &0.54 $  \pm$  0.10&   0.52  \\
HoIX$-$3$-$1168     &   596.137, 284.848 &  09 57 33.7 +69 02 14&  4.0 &  0.3 & $-$6.00 $ \pm$ 0.08 &0.30 $  \pm$  0.11&   0.39  \\
HoIX$-$3$-$1322     &   383.979, 347.673 &  09 57 33.7 +69 02 36&  4.5 &  0.3 & $-$6.00 $ \pm$ 0.08 &0.30 $  \pm$  0.11&   0.14  \\
HoIX$-$3$-$1565     &   376.681, 438.844 &  09 57 32.2 +69 02 39&  4.1 &  0.1 & $-$5.31 $ \pm$ 0.08 &0.74 $  \pm$  0.11&   0.08  \\
HoIX$-$3$-$1664     &   378.637, 471.226 &  09 57 31.6 +69 02 40&  4.1 &  0.0 & $-$6.20 $ \pm$ 0.07 &0.37 $  \pm$  0.10&   0.12  \\
HoIX$-$3$-$1932     &   264.772, 567.834 &  09 57 30.5 +69 02 54&  9.4 &  0.3 & $-$6.61 $ \pm$ 0.07 &0.71 $  \pm$  0.10&   0.38  \\
HoIX$-$3$-$2116     &   376.956, 643.274 &  09 57 28.5 +69 02 45&  5.0 &  0.0 & $-$7.26 $ \pm$ 0.07 &0.37 $  \pm$  0.10&   0.40  \\
HoIX$-$3$-$2129     &   155.863, 649.230 &  09 57 29.7 +69 03 06&  5.1 &  0.1 & $-$5.83 $ \pm$ 0.08 &0.38 $  \pm$  0.11&   0.61  \\
HoIX$-$3$-$2158     &   200.676, 656.790 &  09 57 29.3 +69 03 02&  6.2 &  0.0 & $-$6.98 $ \pm$ 0.08 &0.30 $  \pm$  0.11&   0.57  \\
HoIX$-$3$-$2373     &   558.027, 742.226 &  09 57 25.8 +69 02 31&  7.9 &  0.1 & $-$6.04 $ \pm$ 0.08 &1.31 $  \pm$  0.11&   0.61  \\
HoIX$-$3$-$2376     &   322.360, 744.024 &  09 57 27.1 +69 02 53&  4.8 &  0.0 & $-$5.90 $ \pm$ 0.08 &0.30 $  \pm$  0.11&   0.59  \\
HoIX$-$3$-$2409     &   266.049, 762.700 &  09 57 27.0 +69 02 59&  3.8 &  0.2 & $-$7.39 $ \pm$ 0.07 &0.68 $  \pm$  0.10&   0.66  \\
HoIX$-$4$-$1038     &   172.803, 269.307 &  09 57 40.0 +69 03 25&  5.5 &  0.1 & $-$9.05 $ \pm$ 0.07 &0.44 $  \pm$  0.10&   1.23  \\
HoIX$-$4$-$1085     &   307.374, 301.260 &  09 57 37.8 +69 03 32&  3.8 &  0.1 & $-$5.23 $ \pm$ 0.09 &0.48 $  \pm$  0.13&   1.18  \\
E540-030$-$4$-$1183$^*$ &   150.605, 590.136 &  00 49 20.6 $-$18 02 51&  6.2 &  0.1 & $-$5.37 $ \pm$ 0.08 &1.24 $  \pm$  0.11&   1.65  \\
E294-010$-$3$-$1104 &   485.338, 317.674 &  00 26 32.6 $-$41 51 10&  6.7 &  0.3 & $-$5.32 $ \pm$ 0.07 &1.34 $  \pm$  0.10&   0.14  \\
KK027$-$4$-$721     &   435.879, 190.280 &  03 21 10.0 $-$66 18 26&  7.5 &  0.0 & $-$6.36 $ \pm$ 0.07 &1.15 $  \pm$  0.10&   0.62  \\
Sc22$-$2$-$879      &   108.743, 520.881 &  00 23 53.3 $-$24 41 39& 12.2 &  0.2 & $-$6.11 $ \pm$ 0.08 &1.15 $  \pm$  0.11&   0.88  \\
Sc22$-$2$-$100$^*$  &   734.953, 661.836 &  00 23 55.5 $-$24 40 44&  8.3 &  0.0 & $-$5.90 $ \pm$ 0.07 &0.96 $  \pm$  0.11&   2.17  \\
Sc22$-$4$-$106$^*$  &   598.857, 570.335 &  00 23 44.9 $-$24 42 09&  4.9 &  0.2 & $-$6.05 $ \pm$ 0.07 &0.96 $  \pm$  0.11&   1.86  \\
DDO113$-$2$-$579$^*$&   765.664, 227.122 &  12 14 52.8 +36 14 38&  7.9 &  0.2 & $-$5.60 $ \pm$ 0.07 &1.38 $  \pm$  0.10&   1.54  \\
DDO113$-$4$-$690    &   368.908,  76.646 &  12 14 54.4 +36 12 55&  6.5 &  0.1 & $-$5.27 $ \pm$ 0.08 &0.99 $  \pm$  0.11&   0.61  \\
U7605$-$3$-$1503    &   547.917, 423.740 &  12 28 39.9 +35 43 01& 12.2 &  0.3 & $-$6.44 $ \pm$ 0.08 &1.18 $  \pm$  0.11&   0.38  \\
KK109$-$3$-$1200$^*$&   538.659, 754.664 &  11 47 08.1 +43 40 28&  4.4 &  0.2 & $-$5.87 $ \pm$ 0.08 &0.92 $  \pm$  0.11&   0.95  \\
U7298$-$3$-$1280    &   741.363, 371.228 &  12 16 27.3 +52 13 09&  4.6 &  0.1 & $-$5.67 $ \pm$ 0.08 &0.66 $  \pm$  0.11&   0.88  \\
U8308$-$2$-$1198    &   145.016, 739.639 &  13 13 27.2 +46 19 23&  7.5 &  0.0 & $-$5.52 $ \pm$ 0.08 &1.08 $  \pm$  0.12&   0.90  \\
U8308$-$3$-$2040    &    39.610, 748.843 &  13 13 17.0 +46 19 07&  9.1 &  0.1 & $-$6.62 $ \pm$ 0.08 &1.46 $  \pm$  0.11&   1.21  \\
U8308$-$4$-$893$^*$ &   663.358, 350.654 &  13 13 15.3 +46 19 32&  5.1 &  0.1 & $-$6.30 $ \pm$ 0.07 &1.55 $  \pm$  0.10&   1.59  \\
U8308$-$4$-$971$^*$ &   699.289, 438.755 &  13 13 14.4 +46 19 35&  8.3 &  0.1 & $-$5.62 $ \pm$ 0.09 &1.20 $  \pm$  0.12&   1.78  \\
KK211$-$3$-$917     &   296.748, 173.889 &  13 42 08.0 $-$45 12 29&  6.3 &  0.1 & $-$6.86 $ \pm$ 0.07 &0.91 $  \pm$  0.10&   0.48  \\
KK211$-$3$-$149     &   429.726, 419.984 &  13 42 05.6 $-$45 12 20&  6.1 &  0.2 & $-$7.82 $ \pm$ 0.07 &0.95 $  \pm$  0.10&      0  \\
KK221$-$2$-$608     &   550.374, 168.636 &  13 48 54.9 $-$47 00 10&  5.0 &  0.1 & $-$8.04 $ \pm$ 0.07 &1.00 $  \pm$  0.10&   1.78  \\
KK221$-$2$-$883     &   404.756, 354.161 &  13 48 52.8 $-$47 00 19&  8.3 &  0.1 & $-$7.07 $ \pm$ 0.07 &1.00 $  \pm$  0.10&   1.41  \\
KK221$-$2$-$966     &   140.354, 399.808 &  13 48 50.3 $-$47 00 10&  5.7 &  0.0 & $-$9.80 $ \pm$ 0.07 &0.98 $  \pm$  0.10&   0 90  \\
KK221$-$2$-$1090    &    78.866, 479.706 &  13 48 49.4 $-$47 00 14&  8.7 &  0.0 & $-$7.77 $ \pm$ 0.07 &0.91 $  \pm$  0.10&   0.77  \\
KK221$-$3$-$1062    &   301.010, 266.780 &  13 48 48.2 $-$46 59 46&  9.1 &  0.3 & $-$6.10 $ \pm$ 0.08 &0.93 $  \pm$  0.11&   0.37  \\
KK200$-$3$-$1696    &   785.776, 763.252 &  13 24 32.2 $-$30 58 11&  9.2 &  0.1 & $-$5.68 $ \pm$ 0.09 &1.03 $  \pm$  0.12&   1.17  \\
KK84$-$2$-$785      &   173.139, 447.814 &  10 05 35.8 $-$07 44 08& 11.6 &  0.0 & $-$7.30 $ \pm$ 0.08 &1.00 $  \pm$  0.11&   2.46  \\
KK84$-$2$-$974      &   432.320, 678.779 &  10 05 37.5 $-$07 43 44& 14.9 &  0.0 & $-$6.64 $ \pm$ 0.09 &0.57 $  \pm$  0.14&   3.82  \\
KK84$-$3$-$705      &   460.100, 105.800 &  10 05 35.7 $-$07 44 26&  9.2 &  0.1 & $-$7.45 $ \pm$ 0.08 &1.04 $  \pm$  0.11&   1.66  \\
KK84$-$3$-$830      &   395.303, 454.705 &  10 05 35.1 $-$07 44 60& 10.6 &  0.1 & $-$9.68 $ \pm$ 0.07 &0.96 $  \pm$  0.10&      0  \\
KK84$-$3$-$917      &   624.322, 609.065 &  10 05 36.5 $-$07 45 17& 10.4 &  0.1 & $-$7.52 $ \pm$ 0.08 &1.06 $  \pm$  0.11&   1.29  \\
KK84$-$4$-$666      &   538.305, 207.182 &  10 05 31.5 $-$07 45 04& 10.6 &  0.0 & $-$8.37 $ \pm$ 0.07 &1.26 $  \pm$  0.10&   2.45  \\
KK84$-$4$-$789      &   290.956, 379.188 &  10 05 30.5 $-$07 44 38& 19.4 &  0.3 & $-$7.08 $ \pm$ 0.08 &0.88 $  \pm$  0.12&   3.48  \\
KK84$-$4$-$967      &   571.750, 560.942 &  10 05 29.1 $-$07 45 04& 12.0 &  0.2 & $-$6.81 $ \pm$ 0.08 &0.95 $  \pm$  0.12&   4.10  \\
U9240$-$3$-$4557    &   545.377, 491.651 &  14 24 45.0 +44 31 36&  3.6 &  0.2 & $-$7.22 $ \pm$ 0.07 &0.61 $  \pm$  0.10&   0.19  \\
KK112$-$3$-$976     &   447.781, 261.936 &  11 54 43.3 $-$33 33 44&  9.1 &  0.1 & $-$5.93 $ \pm$ 0.08 &1.25 $  \pm$  0.11&   0.28  \\
KK112$-$4$-$742     &   183.817, 164.666 &  11 54 47.3 $-$33 33 23& 11.8 &  0.1 & $-$6.21 $ \pm$ 0.08 &0.88 $  \pm$  0.11&   1.48  \\
KK112$-$4$-$792     &   444.689, 215.677 &  11 54 46.6 $-$33 32 58& 15.0 &  0.1 & $-$6.77 $ \pm$ 0.08 &1.22 $  \pm$  0.11&   1.43  \\
E490-017$-$3$-$1769 &   306.876, 462.734 &  06 37 57.0 $-$26 00 08&  6.6 &  0.2 & $-$7.06 $ \pm$ 0.08 &0.30 $  \pm$  0.11&   0.05  \\
E490-017$-$3$-$1861 &   246.213, 480.112 &  06 37 57.3 $-$26 00 13&  6.4 &  0.1 & $-$7.38 $ \pm$ 0.07 &0.87 $  \pm$  0.10&   0.16  \\
E490-017$-$3$-$1956 &   507.111, 498.721 &  06 37 57.0 $-$25 59 48&  5.1 &  0.3 & $-$5.37 $ \pm$ 0.09 &0.55 $  \pm$  0.13&   0.37  \\
E490-017$-$3$-$2035 &   374.045, 514.220 &  06 37 57.3 $-$26 00 00&  7.1 &  0.3 & $-$7.30 $ \pm$ 0.07 &0.41 $  \pm$  0.11&   0.12  \\
E490-017$-$3$-$2509 &   657.358, 639.546 &  06 37 57.7 $-$25 59 30&  8.4 &  0.2 & $-$5.69 $ \pm$ 0.08 &1.10 $  \pm$  0.11&   0.75  \\
\noalign{\smallskip}
\hline
\end{tabular}
\end{table*}
\addtocounter{table}{-1}
\begin{table*}
\centering
\caption{--continued.}
\scriptsize
\begin{tabular}{lcclrccl} \\ \hline\hline\noalign{\smallskip}
GCC             &     X, Y           &    RA (J2000) DEC     & $ r_h$&  $e$  & $ V_{0}$           & $(V-I)_{0}$    &   $d_{\rm proj}$ \\
\noalign{\smallskip}
\hline\noalign{\smallskip}
KK065$-$3$-$1095    &   290.890, 434.702 &  07 42 29.4 +16 34 29& 11.5 &  0.1 & $-$6.75 $ \pm$ 0.08 &1.41 $  \pm$  0.11&   0.33  \\
U4115$-$2$-$1042    &   190.123, 708.172 &  07 57 04.9 +14 22 25& 11.9 &  0.3 & $-$6.00 $ \pm$ 0.08 &1.03 $  \pm$  0.12&   1.72  \\
U4115$-$3$-$784     &   614.000, 125.730 &  07 57 03.8 +14 22 43&  9.4 &  0.1 & $-$7.53 $ \pm$ 0.07 &0.93 $  \pm$  0.11&   1.19  \\
U4115$-$4$-$1477    &   607.416, 740.185 &  07 57 04.1 +14 24 58&  8.0 &  0.0 & $-$5.37 $ \pm$ 0.10 &0.91 $  \pm$  0.16&   2.63  \\
UA438$-$3$-$2004    &   299.313, 413.097 &  23 26 28.3 $-$32 23 06&  3.7 &  0.2 & $-$8.67 $ \pm$ 0.07 &0.96 $  \pm$  0.10&   0.16  \\
UA438$-$3$-$3325    &   733.228, 621.063 &  23 26 26.7 $-$32 23 49&  3.7 &  0.1 & $-$5.96 $ \pm$ 0.07 &1.02 $  \pm$  0.10&   0.41  \\
U3755$-$2$-$652     &   378.719, 209.939 &  07 13 50.1 +10 32 15&  5.5 &  0.1 & $-$8.22 $ \pm$ 0.07 &0.98 $  \pm$  0.10&   1.84  \\
U3755$-$2$-$675     &   107.413, 234.823 &  07 13 50.4 +10 31 48&  8.1 &  0.1 & $-$5.75 $ \pm$ 0.09 &1.19 $  \pm$  0.12&   1.21  \\
U3755$-$2$-$863     &    61.125, 358.956 &  07 13 51.2 +10 31 44&  5.2 &  0.1 & $-$6.93 $ \pm$ 0.08 &1.03 $  \pm$  0.11&   0.96  \\
U3755$-$3$-$727     &   494.971,  57.246 &  07 13 52.1 +10 31 43&  9.5 &  0.1 & $-$7.21 $ \pm$ 0.10 &0.60 $  \pm$  0.15&   0.85  \\
U3755$-$3$-$739     &   467.346,  64.840 &  07 13 51.9 +10 31 42&  5.7 &  0.2 & $-$8.67 $ \pm$ 0.07 &0.85 $  \pm$  0.10&   0.83  \\
U3755$-$3$-$754     &   380.674,  72.749 &  07 13 51.3 +10 31 41&  8.1 &  0.0 & $-$6.47 $ \pm$ 0.09 &0.56 $  \pm$  0.14&   0.82  \\
U3755$-$3$-$768     &   404.724,  81.923 &  07 13 51.5 +10 31 40&  5.7 &  0.2 & $-$5.42 $ \pm$ 0.10 &1.05 $  \pm$  0.17&   0.79  \\
U3755$-$3$-$914     &   379.331, 139.121 &  07 13 51.3 +10 31 34&  9.5 &  0.0 & $-$7.60 $ \pm$ 0.08 &0.64 $  \pm$  0.11&   0.65  \\
U3755$-$3$-$974     &   434.698, 161.008 &  07 13 51.7 +10 31 32&  7.0 &  0.1 & $-$5.70 $ \pm$ 0.01 &0.53 $  \pm$  0.16&   0.58  \\
U3755$-$3$-$1045    &   390.476, 186.888 &  07 13 51.4 +10 31 29&  8.3 &  0.3 & $-$6.01 $ \pm$ 0.10 &0.45 $  \pm$  0.16&   0.53  \\
U3755$-$3$-$1182    &   393.821, 228.672 &  07 13 51.5 +10 31 25&  7.9 &  0.2 & $-$8.54 $ \pm$ 0.07 &0.56 $  \pm$  0.10&   0.43  \\
U3755$-$3$-$1256    &   483.184, 249.602 &  07 13 52.1 +10 31 24&  8.6 &  0.0 & $-$7.77 $ \pm$ 0.07 &0.97 $  \pm$  0.10&   0.37  \\
U3755$-$3$-$1257    &   511.799, 250.293 &  07 13 52.3 +10 31 24&  6.2 &  0.0 & $-$8.71 $ \pm$ 0.07 &0.97 $  \pm$  0.10&   0.40  \\
U3755$-$3$-$1364    &   354.699, 277.022 &  07 13 51.2 +10 31 20&  6.0 &  0.1 & $-$7.00 $ \pm$ 0.08 &0.49 $  \pm$  0.12&   0.36  \\
U3755$-$3$-$1611    &   318.852, 334.848 &  07 13 51.0 +10 31 14&  7.5 &  0.0 & $-$7.48 $ \pm$ 0.08 &0.83 $  \pm$  0.11&   0.34  \\
U3755$-$3$-$1616    &   615.652, 335.551 &  07 13 53.0 +10 31 16&  7.5 &  0.2 & $-$6.40 $ \pm$ 0.09 &0.69 $  \pm$  0.12&   0.46  \\
U3755$-$3$-$1732    &   195.199, 365.183 &  07 13 50.2 +10 31 11& 10.0 &  0.1 & $-$6.09 $ \pm$ 0.08 &1.01 $  \pm$  0.11&   0.62  \\
U3755$-$3$-$1737    &   466.051, 366.036 &  07 13 52.0 +10 31 12&  8.3 &  0.2 & $-$6.71 $ \pm$ 0.08 &1.05 $  \pm$  0.11&   0.09  \\
U3755$-$3$-$1963    &   453.046, 423.153 &  07 13 51.9 +10 31 06&  9.1 &  0.1 & $-$6.46 $ \pm$ 0.09 &0.56 $  \pm$  0.14&   0.09  \\
U3755$-$3$-$2027    &   369.997, 444.823 &  07 13 51.4 +10 31 04&  6.3 &  0.2 & $-$5.80 $ \pm$ 0.10 &0.60 $  \pm$  0.15&   0.22  \\
U3755$-$3$-$2123    &   530.273, 487.378 &  07 13 52.5 +10 31 00&  6.5 &  0.1 & $-$7.76 $ \pm$ 0.07 &0.42 $  \pm$  0.11&   0.33  \\
U3755$-$3$-$2168    &   363.214, 511.255 &  07 13 51.4 +10 30 57&  8.7 &  0.2 & $-$7.63 $ \pm$ 0.07 &1.00 $  \pm$  0.10&   0.36  \\
U3755$-$3$-$2204    &   496.118, 528.265 &  07 13 52.3 +10 30 56&  6.6 &  0.1 & $-$6.07 $ \pm$ 0.09 &0.82 $  \pm$  0.13&   0.37  \\
U3755$-$3$-$2334    &   637.259, 611.302 &  07 13 53.3 +10 30 49&  6.7 &  0.1 & $-$6.90 $ \pm$ 0.07 &0.51 $  \pm$  0.11&   0.74  \\
U3755$-$3$-$2363    &   471.076, 639.739 &  07 13 52.2 +10 30 45&  8.3 &  0.0 & $-$7.75 $ \pm$ 0.07 &0.49 $  \pm$  0.10&   0.63  \\
U3755$-$3$-$2368    &   526.825, 646.637 &  07 13 52.5 +10 30 45&  7.0 &  0.3 & $-$7.44 $ \pm$ 0.07 &0.53 $  \pm$  0.11&   0.68  \\
U3755$-$3$-$2398    &   555.514, 670.612 &  07 13 52.7 +10 30 43&  8.5 &  0.3 & $-$6.17 $ \pm$ 0.09 &1.13 $  \pm$  0.13&   0.76  \\
U3755$-$3$-$2401    &   392.910, 676.091 &  07 13 51.6 +10 30 41& 12.0 &  0.0 & $-$7.54 $ \pm$ 0.07 &0.97 $  \pm$  0.10&   0.73  \\
U3755$-$3$-$2403    &   231.105, 677.272 &  07 13 50.5 +10 30 40&  6.8 &  0.0 & $-$6.79 $ \pm$ 0.07 &0.92 $  \pm$  0.10&   0.89  \\
U3755$-$3$-$2408    &   496.303, 682.668 &  07 13 52.3 +10 30 41&  8.6 &  0.0 & $-$6.89 $ \pm$ 0.08 &0.43 $  \pm$  0.11&   0.75  \\
U3755$-$3$-$2459    &   457.060, 743.123 &  07 13 52.1 +10 30 35&  8.3 &  0.0 & $-$7.98 $ \pm$ 0.07 &0.57 $  \pm$  0.10&   0.89  \\
U3755$-$4$-$566     &   188.706,  59.303 &  07 13 48.9 +10 31 27&  8.6 &  0.3 & $-$6.25 $ \pm$ 0.09 &0.87 $  \pm$  0.14&   1.31  \\
\noalign{\smallskip}
\hline
\end{tabular}
\end{table*}

\begin{table*}
\centering
\caption{The King law approximation parameters for globular cluster candidates in
nearby LSB dwarf galaxies. The table contains the following columns:
identifier of each cluster (as in Table~\ref{tab:smpl1}), reddening
corrected $V$-band central surface brightness in
mag/arcsec$^2$ and corresponding error, reddening corrected 
 $I$-band central surface brightness in mag/arcsec$^2$ and
corresponding error, King core radius $r_c$ and corresponding error, King
tidal radius $r_t$ and corresponding error, and the King concentration
parameter $c\!=\!r_t/r_c$.}
\label{tab:smpl2}
\scriptsize
\begin{tabular}{lcclll} \\ \hline\hline\noalign{\smallskip}
GCC             & $ \mu_{V,0}  $            & $ \mu_{I,0} $    & $ r_c $               & $ r_t $  & $c$    \\
\noalign{\smallskip}\hline\noalign{\smallskip}
DDO53$-$3$-$1120    & 20.92 $  \pm $  0.08 & 19.67 $ \pm $  0.01 & 4.16 $  \pm $    0.09 & 21.3 $  \pm  $   1.0 &   5.2 \\
BK3N$-$2$-$863$^*$  & 20.95 $  \pm $  0.13 & 20.70 $ \pm $  0.03 & 2.38 $  \pm $    0.38 & 26.4 $  \pm  $   4.5 &  11.1 \\
KDG73$-$2$-$378$^*$ & 21.33 $  \pm $  0.05 & 20.09 $ \pm $  0.03 & 3.47 $  \pm $    0.21 & 124.:                &  36.0 \\
KK77$-$4$-$939      & 20.64 $  \pm $  0.68 & 20.44 $ \pm $  0.07 & 2.26 $  \pm $    0.22 & 17.1 $  \pm  $   3.4 &   7.6 \\
KK77$-$4$-$1162     & 21.03 $  \pm $  0.04 & 20.35 $ \pm $  0.05 & 2.78 $  \pm $    0.18 & 18.0 $  \pm  $   2.4 &   6.5 \\
KK77$-$4$-$1165     & 21.42 $  \pm $  0.05 & 20.46 $ \pm $  0.17 & 3.19 $  \pm $    0.22 & 15.9 $  \pm  $   0.7 &   5.0 \\
KDG61$-$3$-$1325    & 18.45 $  \pm $  0.03 & 17.59 $ \pm $  0.05 & 1.88 $  \pm $    0.06 & 33.8 $  \pm  $   2.7 &  18.0 \\
KDG63$-$3$-$1168    & 18.93 $  \pm $  0.04 & 18.14 $ \pm $  0.03 & 1.90 $  \pm $    0.06 & 53.4 $  \pm  $  11.2 &  28.1 \\
DDO78$-$1$-$167     & 18.69 $  \pm $  0.11 & 17.75 $ \pm $  0.03 & 3.53 $  \pm $    0.14 & 46.7 $  \pm  $   6.8 &  13.2 \\
DDO78$-$3$-$1082    & 17.93 $  \pm $  0.04 & 16.98 $ \pm $  0.02 & 2.89 $  \pm $    0.06 & 34.6 $  \pm  $   2.1 &  12.0 \\
BK6N$-$2$-$524      & 20.40 $  \pm $  0.13 & 20.04 $ \pm $  0.05 & 1.54 $  \pm $    0.15 & 90.2 $  \pm  $  56.7 &  58.7 \\
BK6N$-$4$-$789      & 20.47 $  \pm $  0.11 & 19.66 $ \pm $  0.01 & 1.91 $  \pm $    0.18 & 68.0 $  \pm  $   8.1 &  34.9 \\
Garland$-$1$-$728   & 16.76 $  \pm $  0.04 & 15.90 $ \pm $  0.08 & 0.69 $  \pm $    0.00 & 42.3 $  \pm  $  14.6 &  61.0 \\
HoIX$-$3$-$866      & 17.88 $  \pm $  0.07 & 18.14 $ \pm $  0.11 & 1.51 $  \pm $    0.06 & 42.3 $  \pm  $   7.1 &  28.0 \\
HoIX$-$3$-$1168     & 19.69 $  \pm $  0.07 & 19.71 $ \pm $  0.42 & 2.00 $  \pm $    0.15 & 12.3 $  \pm  $   1.1 &   6.2 \\
HoIX$-$3$-$1322     & 20.02 $  \pm $  0.01 & 20.37 $ \pm $  0.08 & 2.34 $  \pm $    0.00 & 18.0 $  \pm  $   0.5 &   7.7 \\
HoIX$-$3$-$1565     & 19.34 $  \pm $  0.31 & 19.10 $ \pm $  0.64 & 1.06 $  \pm $    0.44 & 16.8 $  \pm  $   7.4 &  13.5 \\
HoIX$-$3$-$1664     & 19.28 $  \pm $  0.11 & 19.17 $ \pm $  0.39 & 1.18 $  \pm $    0.09 & 116. $  \pm  $  87.2 &  98.0 \\
HoIX$-$3$-$1932     & 20.06 $  \pm $  0.03 & 19.60 $ \pm $  0.07 & 3.50 $  \pm $    0.15 & 23.0 $  \pm  $   2.6 &   6.6 \\
HoIX$-$3$-$2116     & 18.53 $  \pm $  0.19 & 18.65 $ \pm $  0.06 & 2.18 $  \pm $    0.14 & 21.4 $  \pm  $   3.0 &   8.7 \\
HoIX$-$3$-$2129     & 20.42 $  \pm $  0.17 & 20.30 $ \pm $  0.21 & 2.09 $  \pm $    0.45 & 975.:                &  466.1\\
HoIX$-$3$-$2158     & 19.41 $  \pm $  0.04 & 19.44 $ \pm $  0.07 & 3.69 $  \pm $    0.28 & 11.8 $  \pm  $   0.9 &   3.2 \\
HoIX$-$3$-$2373     & 21.11 $  \pm $  0.03 & 19.70 $ \pm $  0.06 & 3.17 $  \pm $    0.31 & 22.3 $  \pm  $   2.6 &   7.1 \\
HoIX$-$3$-$2376     & 19.60 $  \pm $  0.30 & 20.34 $ \pm $  0.16 & 1.46 $  \pm $    0.46 & 76.3:                &  52.4 \\
HoIX$-$3$-$2409     & 18.31 $  \pm $  0.12 & 17.91 $ \pm $  0.03 & 1.91 $  \pm $    0.06 & 14.4 $  \pm  $   0.5 &   7.5 \\
HoIX$-$4$-$1038     & 17.06 $  \pm $  0.10 & 16.81 $ \pm $  0.11 & 2.08 $  \pm $    0.19 & 27.1 $  \pm  $   3.9 &  13.1 \\
HoIX$-$4$-$1085     & 20.24 $  \pm $  0.27 & 20.50 $ \pm $  0.40 & 1.40 $  \pm $    0.29 & 18.3 $  \pm  $   4.8 &  13.2 \\
E540-030$-$4$-$1183$^*$ &21.13 $  \pm $  0.02 & 20.15 $ \pm $  0.03 & 2.92 $  \pm $    0.09 & 28.3 $  \pm  $   2.9 &   9.7 \\
E294-010$-$3$-$1104 & 20.96 $  \pm $  0.05 & 19.52 $ \pm $  0.06 & 3.64 $  \pm $    0.50 & 17.0 $  \pm  $   3.3 &   4.7 \\
KK027$-$4$-$721     & 20.39 $  \pm $  0.09 & 19.55 $ \pm $  0.03 & 3.33 $  \pm $    0.11 & 41.1 $  \pm  $   4.9 &  12.3 \\
Scu22$-$2$-$879     & 22.08 $  \pm $  0.04 & 20.77 $ \pm $  0.05 & 6.06 $  \pm $    0.62 & 51.3 $  \pm  $  16.9 &   8.5 \\
Scu22$-$2$-$100$^*$ & 20.72 $  \pm $  0.04 & 20.29 $ \pm $  0.03 & 3.44 $  \pm $    0.20 & 22.2 $  \pm  $   2.2 &   6.5 \\
Scu22$-$4$-$106$^*$ & 20.12 $  \pm $  0.10 & 19.44 $ \pm $  0.07 & 1.91 $  \pm $    0.18 & 33.4 $  \pm  $   7.0 &  17.5 \\
DDO113$-$2$-$579$^*$ & 21.26 $  \pm $  0.05 & 19.53 $ \pm $  0.03 & 2.73 $  \pm $    0.14 & 14.1 $  \pm  $   1.3 &   5.2 \\
DDO113$-$4$-$690     & 21.32 $  \pm $  0.02 & 20.18 $ \pm $  0.07 & 3.74 $  \pm $    0.19 & 19.4 $  \pm  $   2.4 &   5.2 \\
U7605$-$3$-$1503    & 21.49 $  \pm $  0.07 & 20.48 $ \pm $  0.05 & 6.16 $  \pm $    0.71 & 80.3 $  \pm  $  79.7 &  13.1 \\
KK109$-$3$-$1200$^*$& 19.94 $  \pm $  0.17 & 19.40 $ \pm $  0.20 & 1.70 $  \pm $    0.22 & 19.2 $  \pm  $   3.0 &  11.3 \\
U7298$-$3$-$1280    & 20.56 $  \pm $  0.16 & 19.88 $ \pm $  0.45 & 2.42 $  \pm $    0.45 & 16.1 $  \pm  $   4.6 &   6.7 \\
U8308$-$2$-$1198    & 21.61 $  \pm $  0.03 & 20.17 $ \pm $  0.06 & 2.78 $  \pm $    0.20 & 27.4 $  \pm  $   3.7 &   9.9 \\
U8308$-$3$-$2040    & 20.62 $  \pm $  0.08 & 19.07 $ \pm $  0.03 & 3.25 $  \pm $    0.17 & 34.2 $  \pm  $  15.2 &  10.5 \\
U8308$-$4$-$893$^*$ & 20.03 $  \pm $  0.02 & 18.54 $ \pm $  0.06 & 1.91 $  \pm $    0.00 & 66.9 $  \pm  $   5.8 &  35.0 \\
U8308$-$4$-$971$^*$ & 21.83 $  \pm $  0.06 & 20.64 $ \pm $  0.02 & 3.68 $  \pm $    0.30 & 132. $  \pm  $  50.0 &  36.2 \\
KK211$-$3$-$917     & 19.55 $  \pm $  0.04 & 18.98 $ \pm $  0.03 & 2.54 $  \pm $    0.12 & 51.9 $  \pm  $  13.9 &  20.5 \\
KK211$-$3$-$149     & 18.99 $  \pm $  0.03 & 18.18 $ \pm $  0.05 & 2.98 $  \pm $    0.12 & 113. $  \pm  $  51.9 &  38.1 \\
KK221$-$2$-$608     & 18.03 $  \pm $  0.03 & 17.50 $ \pm $  0.08 & 1.78 $  \pm $    0.06 & 49.4 $  \pm  $   3.9 &  27.8 \\
KK221$-$2$-$883     & 20.04 $  \pm $  0.03 & 19.30 $ \pm $  0.04 & 3.43 $  \pm $    0.17 & 71.4 $  \pm  $  28.8 &  20.8 \\
KK221$-$2$-$966     & 16.57 $  \pm $  0.04 & 15.90 $ \pm $  0.06 & 2.26 $  \pm $    0.11 & 46.0 $  \pm  $   6.2 &  20.4 \\
KK221$-$2$-$1090    & 19.17 $  \pm $  0.04 & 18.58 $ \pm $  0.02 & 3.57 $  \pm $    0.09 & 30.6 $  \pm  $   3.3 &   8.6 \\
KK221$-$3$-$1062    & 21.77 $  \pm $  0.06 & 21.00 $ \pm $  0.06 & 5.85 $  \pm $    0.85 & 43.1 $  \pm  $  21.5 &   7.4 \\
KK200$-$3$-$1696    & 21.48 $  \pm $  0.06 & 20.87 $ \pm $  0.04 & 4.25 $  \pm $    0.42 & 38.0 $  \pm  $  13.2 &   9.0 \\
KK84$-$2$-$785      & 20.25 $  \pm $  0.12 & 19.69 $ \pm $  0.13 & 4.54 $  \pm $    0.50 & 39.9 $  \pm  $   5.3 &   8.8 \\
KK84$-$2$-$974      & 21.48 $  \pm $  0.10 & 21.08 $ \pm $  0.04 & 5.31 $  \pm $    0.61 & 38.5 $  \pm  $   3.9 &   7.3 \\
KK84$-$3$-$705      & 21.41 $  \pm $  0.33 & 20.03 $ \pm $  0.10 & 10.6 $  \pm $    1.86 & 36.8 $  \pm  $  14.1 &   3.4 \\
KK84$-$3$-$830      & 18.19 $  \pm $  0.03 & 17.16 $ \pm $  0.10 & 3.19 $  \pm $    0.18 & 90.4 $  \pm  $   4.0 &  28.4 \\
KK84$-$3$-$917      & 20.56 $  \pm $  0.10 & 19.62 $ \pm $  0.17 & 4.84 $  \pm $    0.43 & 40.9 $  \pm  $   4.7 &   8.5 \\
KK84$-$4$-$666      & 19.32 $  \pm $  0.07 & 18.44 $ \pm $  0.03 & 3.81 $  \pm $    0.18 & 45.3 $  \pm  $   1.8 &  11.9 \\
KK84$-$4$-$789      & 21.59 $  \pm $  0.14 & 20.94 $ \pm $  0.10 & 6.85 $  \pm $    1.56 & 57.7 $  \pm  $  34.1 &   8.4 \\
KK84$-$4$-$967      & 21.50 $  \pm $  0.05 & 20.61 $ \pm $  0.14 & 9.90 $  \pm $    1.12 & 22.2 $  \pm  $   1.7 &   2.3 \\
U9240$-$3$-$4557    & 18.37 $  \pm $  0.01 & 17.64 $ \pm $  0.10 & 1.81 $  \pm $    0.00 & 20.2 $  \pm  $   0.8 &  11.2 \\
KK112$-$3$-$976     & 21.03 $  \pm $  0.03 & 20.28 $ \pm $  0.05 & 3.37 $  \pm $    0.12 & 40.6 $  \pm  $   5.3 &  12.1 \\
KK112$-$4$-$742     & 21.40 $  \pm $  0.06 & 20.76 $ \pm $  0.05 & 4.64 $  \pm $    0.41 & 68.8 $  \pm  $  38.5 &  14.8 \\
KK112$-$4$-$792     & 21.26 $  \pm $  0.03 & 19.99 $ \pm $  0.02 & 6.20 $  \pm $    0.20 & 41.9 $  \pm  $   3.6 &   6.8 \\
E490-017$-$3$-$1769 & 19.49 $  \pm $  0.03 & 19.33 $ \pm $  0.10 & 3.16 $  \pm $    0.15 & 38.6 $  \pm  $   6.9 &  12.2 \\
E490-017$-$3$-$1861 & 18.86 $  \pm $  0.08 & 18.24 $ \pm $  0.04 & 2.76 $  \pm $    0.12 & 28.8 $  \pm  $   3.8 &  10.5 \\
E490-017$-$3$-$1956 & 20.85 $  \pm $  0.28 & 20.59 $ \pm $  0.01 & 2.74 $  \pm $    0.09 & 22.8 $  \pm  $  26.9 &   8.3 \\
E490-017$-$3$-$2035 & 19.23 $  \pm $  0.03 & 18.80 $ \pm $  0.06 & 3.27 $  \pm $    0.18 & 38.5 $  \pm  $  15.2 &  11.8 \\
E490-017$-$3$-$2509 & 21.00 $  \pm $  0.01 & 20.38 $ \pm $  0.04 & 4.79 $  \pm $    0.09 & 16.5 $  \pm  $   0.4 &   3.5 \\
\noalign{\smallskip}\hline
\end{tabular}
\end{table*}
\addtocounter{table}{-1}
\begin{table*}
\centering
\caption{-- continued.}
\scriptsize
\begin{tabular}{lcclll} \\ \hline\hline\noalign{\smallskip}
GCC             & $ \mu_{V0}  $            & $ \mu_{I0} $            & $ r_c $               & $ r_t $ & $c$     \\
\noalign{\smallskip}\hline\noalign{\smallskip}
KK065$-$3$-$1095    & 20.96 $  \pm $  0.03 & 19.50 $ \pm $  0.01 & 5.28 $  \pm $    0.12 & 55.3 $  \pm  $   6.0 &  10.5 \\
U4115$-$2$-$1042    & 21.74 $  \pm $  0.08 & 21.07 $ \pm $  0.07 & 4.89 $  \pm $    0.71 & 293.:                      &  59.9 \\
U4115$-$3$-$784     & 19.82 $  \pm $  0.14 & 19.21 $ \pm $  0.09 & 3.11 $  \pm $    0.48 & 64.6 $  \pm  $  38.7 &  20.8 \\
U4115$-$4$-$1477    & 21.74 $  \pm $  0.36 & 20.96 $ \pm $  0.05 & 4.23 $  \pm $    0.38 & 25.7 $  \pm  $   5.1 &   6.1 \\
UA438$-$3$-$2004    & 17.05 $  \pm $  0.01 & 16.10 $ \pm $  0.04 & 1.73 $  \pm $    0.04 & 33.3 $  \pm  $   2.6 &  19.3 \\
UA438$-$3$-$3325    & 19.42 $  \pm $  0.08 & 18.62 $ \pm $  0.03 & 1.67 $  \pm $    0.06 & 31.2 $  \pm  $  19.8 &  18.7 \\
U3755$-$2$-$652     & 17.78 $  \pm $  0.06 & 17.19 $ \pm $  0.03 & 1.78 $  \pm $    0.27 & 33.5 $  \pm  $   7.7 & 18.9  \\
U3755$-$2$-$675     & 21.20 $  \pm $  0.05 & 20.24 $ \pm $  0.34 & 4.36 $  \pm $    0.38 & 18.6 $  \pm  $   1.9 &  4.3  \\
U3755$-$2$-$863     & 19.11 $  \pm $  0.15 & 18.92 $ \pm $  0.02 & 1.77 $  \pm $    0.20 & 42.6 $  \pm  $   2.8 & 24.1  \\
U3755$-$3$-$727     & 20.19 $  \pm $  0.08 & 19.82 $ \pm $  0.03 & 4.04 $  \pm $    0.20 & 141. $  \pm  $  97.5 & 35.1  \\
U3755$-$3$-$739     & 18.00 $  \pm $  0.09 & 17.37 $ \pm $  0.03 & 2.30 $  \pm $    0.09 & 67.8 $  \pm  $   7.6 & 29.5  \\
U3755$-$3$-$754     & 20.34 $  \pm $  0.05 & 20.20 $ \pm $  0.09 & 3.93 $  \pm $    0.51 & 25.0 $  \pm  $   6.1 &  6.4  \\
U3755$-$3$-$768     & 20.98 $  \pm $  0.21 & 21.08 $ \pm $  0.25 & 2.25 $  \pm $    0.41 & 38.0 $  \pm  $  15.7 & 17.0  \\
U3755$-$3$-$914     & 19.84 $  \pm $  0.04 & 20.06 $ \pm $  0.03 & 3.43 $  \pm $    0.18 & 94.1 $  \pm  $  46.2 & 27.5  \\
U3755$-$3$-$974     & 21.22 $  \pm $  0.04 & 20.82 $ \pm $  0.03 & 3.47 $  \pm $    0.20 & 15.7 $  \pm  $   5.5 &  4.5  \\
U3755$-$3$-$1045    & 21.30 $  \pm $  0.08 & 21.02 $ \pm $  0.09 & 5.63 $  \pm $    1.41 & 17.5 $  \pm  $   4.8 &  3.1  \\
U3755$-$3$-$1182    & 18.39 $  \pm $  0.02 & 18.06 $ \pm $  0.02 & 3.54 $  \pm $    0.09 & 33.1 $  \pm  $   1.9 &  9.4  \\
U3755$-$3$-$1256    & 19.49 $  \pm $  0.05 & 18.82 $ \pm $  0.01 & 4.13 $  \pm $    0.09 & 36.6 $  \pm  $   1.5 &  8.9  \\
U3755$-$3$-$1257    & 17.74 $  \pm $  0.03 & 17.24 $ \pm $  0.03 & 2.15 $  \pm $    0.09 & 35.4 $  \pm  $   1.8 & 16.5  \\
U3755$-$3$-$1364    & 19.49 $  \pm $  0.22 & 19.52 $ \pm $  0.08 & 2.73 $  \pm $    0.27 & 283.:                & 103.9 \\
U3755$-$3$-$1611    & 19.29 $  \pm $  0.05 & 18.82 $ \pm $  0.06 & 2.70 $  \pm $    0.12 & 38.3 $  \pm  $   4.9 & 14.2  \\
U3755$-$3$-$1616    & 20.83 $  \pm $  0.12 & 20.39 $ \pm $  0.07 & 3.12 $  \pm $    0.48 & 60.7 $  \pm  $  32.7 & 19.5  \\
U3755$-$3$-$1732    & 21.04 $  \pm $  0.08 & 20.37 $ \pm $  0.08 & 3.89 $  \pm $    0.50 & 60.9 $  \pm  $  48.9 & 15.7  \\
U3755$-$3$-$1737    & 20.17 $  \pm $  0.04 & 19.95 $ \pm $  0.05 & 3.02 $  \pm $    0.15 & 34.6 $  \pm  $   4.5 & 11.5  \\
U3755$-$3$-$1963    & 20.91 $  \pm $  0.12 & 20.70 $ \pm $  0.10 & 4.50 $  \pm $    1.03 & 90.4:                & 20.1  \\
U3755$-$3$-$2027    & 20.85 $  \pm $  0.04 & 20.98 $ \pm $  0.05 & 3.53 $  \pm $    0.20 & 16.6 $  \pm  $   1.0 &  4.7  \\
U3755$-$3$-$2123    & 19.22 $  \pm $  0.05 & 19.18 $ \pm $  0.11 & 2.62 $  \pm $    0.12 & 344.:                & 131.4 \\
U3755$-$3$-$2168    & 19.54 $  \pm $  0.03 & 18.93 $ \pm $  0.03 & 4.18 $  \pm $    0.18 & 27.1 $  \pm  $   2.0 &  6.5  \\
U3755$-$3$-$2204    & 20.81 $  \pm $  0.15 & 20.27 $ \pm $  0.04 & 2.98 $  \pm $    0.12 & 19.8 $  \pm  $   6.4 &  6.7  \\
U3755$-$3$-$2334    & 19.90 $  \pm $  0.05 & 19.57 $ \pm $  0.05 & 3.34 $  \pm $    0.22 & 20.1 $  \pm  $   2.0 &  6.0  \\
U3755$-$3$-$2363    & 19.05 $  \pm $  0.08 & 18.86 $ \pm $  0.07 & 3.00 $  \pm $    0.27 & 39.7 $  \pm  $   9.4 & 13.3  \\
U3755$-$3$-$2368    & 19.82 $  \pm $  0.02 & 19.59 $ \pm $  0.01 & 3.35 $  \pm $    0.09 & 33.4 $  \pm  $   2.0 & 10.0  \\
U3755$-$3$-$2398    & 21.39 $  \pm $  0.03 & 20.77 $ \pm $  0.06 & 7.04 $  \pm $    0.48 & 18.0 $  \pm  $   1.1 &  2.6  \\
U3755$-$3$-$2401    & 19.92 $  \pm $  0.04 & 19.63 $ \pm $  0.02 & 4.07 $  \pm $    0.22 & 42.0 $  \pm  $   3.0 & 10.3  \\
U3755$-$3$-$2403    & 20.00 $  \pm $  0.06 & 19.11 $ \pm $  0.10 & 3.20 $  \pm $    0.22 & 20.0 $  \pm  $   2.0 &  6.3  \\
U3755$-$3$-$2408    & 20.21 $  \pm $  0.06 & 20.12 $ \pm $  0.05 & 3.59 $  \pm $    0.31 & 66.8 $  \pm  $  36.7 & 18.6  \\
U3755$-$3$-$2459    & 18.65 $  \pm $  0.04 & 18.74 $ \pm $  0.05 & 2.62 $  \pm $    0.09 & 49.0 $  \pm  $   5.1 & 18.7  \\
U3755$-$4$-$566     & 21.23 $  \pm $  0.03 & 20.59 $ \pm $  0.06 & 4.87 $  \pm $    0.33 & 53.6 $  \pm  $  19.4 & 11.0  \\
\noalign{\smallskip}
\hline
\end{tabular}
\end{table*}

{}

\end{document}